\documentclass[]{ptptex}
\usepackage{times}
\usepackage{amsmath}
\usepackage{amssymb}
\usepackage{bm}

\begin{document}
\title{Adiabatic Continuation between\\ Resonating-Valence-Bond Electron and Spin Liquids}
%
\author{Fusayoshi J. Ohkawa\footnote{E-mail: fohkawa@phys.sci.hokudai.ac.jp}}
\inst{Division of Physics, Hokkaido University, Sapporo 060-0810, Japan}
\date{Received October 29, 2012}
\abst{    
The Hubbard model in the strong-coupling regime is studied by the Kondo-lattice theory.
If no symmetry is broken at a sufficiently low temperature in sufficiently low dimensions,
an electron liquid is stabilized by the Fock-type exchange effect of the superexchange interaction.
Since the stabilization mechanism is none other than the resonating-valence-bond (RVB) mechanism, the electron liquid is none other than an RVB electron liquid.
The RVB electron liquid in the Hubbard model is adiabatically connected to the RVB spin liquid in the Heisenberg model.
}
\PTPindex{I45, I46, I71, I72, I97}

\maketitle

\section{Introduction}
\label{SecIntroduction}
The concept of adiabatic or analytic continuation is the counter concept of symmetry breaking: 
If no symmetry is broken or restored when a phase transforms into another phase as a function of an adiabatic parameter or the temperature $T$, the two phases are adiabatically or analytically continued or connected to each other.
The concept is useful to understand, for example, the Kondo effect. 

The $s$-$d$ model is derived from the Anderson model in the $s$-$d$ limit.
Since the $s$-$d$ exchange interaction is antiferromagnetic,
the ground state of the $s$-$d$ model is a singlet because of the quenching of the localized spin by conduction electrons \cite{yosida,poorman,wilsonKG}. 
The Kondo temperature $T_{\rm K}$ or $k_{\rm B}T_{\rm K}$ is the energy scale of local quantum spin fluctuations.
The phase transformation between a local-moment state at $T\gg T_{\rm K}$ and a local spin liquid at $T\ll T_{\rm K}$ is a crossover, and
no symmetry is broken in the lower temperature phase. Thus, the local-moment state and the local spin liquid are analytically connected to each other \cite{wilsonKG}.
Furthermore, the local spin liquid can be described as the local normal Fermi liquid \cite{nozieres}.
Since local charges carried by the localized spin cannot fluctuate in the least, {\it the existence of charge fluctuations is not a necessary condition for the normal Fermi liquid to be stabilized}.

Since electron correlations are local in the Anderson model, no cooperative effect is possible and no symmetry can be broken in it; therefore,
the conventional perturbative analysis in terms of the onsite repulsion $U$ is valid, no anomalous term appears in the perturbation, and 
the ground state is the local normal Fermi liquid for any finite $U\ge 0$, even in the $s$-$d$ limit \cite{yamada1,yamada2}. Thus, the local Fermi liquid for any finite $U>0$ is adiabatically connected to that for $U=0$. 
The Fermi-liquid relation for the Anderson model in the $s$-$d$ limit is exactly the same as that for the $s$-$d$ model \cite{wilsonKG,nozieres,yamada1,yamada2,shiba}. 
The facts discussed above mean that the normal Fermi liquid in the Anderson model for any $U\ge 0$ is adiabatically connected to the spin liquid in the $s$-$d$ model, {\it even if local charges can fluctuate in the Fermi liquid in the Anderson model but they cannot, in the least, in the spin liquid in the $s$-$d$ model}.
On the basis of the adiabatic continuation, 
it is reasonable that physical properties resemble each other between the Anderson and $s$-$d$ models.

The relation between the Hubbard and Heisenberg models is similar to that between the Anderson and $s$-$d$ models.
%
If $U/|t|\gg 1$ in the half-filled Hubbard model, where $U$ is the onsite repulsion and $-t$ is the transfer integral between nearest neighbors, the band splits into the upper and lower Hubbard bands, which are empty and  full, respectively \cite{hubbard1,hubbard3}; and a narrow midband predicted by Gutzwiller's theory \cite{Gutzwiller1,Gutzwiller2,Gutzwiller3} can appear within the Hubbard gap between them \cite{midband}.
On the other hand, almost all unit cells are singly occupied and the probabilities of empty and double occupancies are almost vanishing but still nonzero.
The superexchange interaction arises from the virtual process allowing empty and double occupancies \cite{Js-mech-pert}.
In field theory, on the other hand, any mutual interaction arises from the virtual exchange of a boson or a bosonic excitation; e.g., the nuclear force arises from that of a Yukawa's meson or a pion.
The virtual process from which the superexchange interaction arises is none other than the virtual exchange of a pair excitation of an electron in the upper Hubbard band and a hole in the lower Hubbard band.
The pair excitation is a bosonic excitation.
The superexchange interaction arises from the virtual exchange of the bosonic excitation \cite{FJO-supJ}.
The superexchange interaction is antiferromagnetic and 
the exchange interaction constant between nearest-neighbor spins is $J=-4t^2/U$.
Thus, the Heisenberg model is derived from the Hubbard model in the Heisenberg limit. 
It is anticipated that physical properties resemble each other between the half-filled Hubbard model with $U/|t|\gg 1$ and the Heisenberg model with $J=-4t^2/U$, except for physical properties directly related to charge fluctuations such as the electrical conductivity.

The role of the superexchange interaction is dual: the cause and suppression of magnetic instability.
Since the superexchange interaction is antiferromagnetic, it is possible for an antiferromagnetic state to be stabilized below the N\'{e}el temperature $T_{\rm N}$.
On the other hand, Fazekas and Anderson proposed the resonating-valence-bond (RVB) theory for the Heisenberg model on the triangular lattice \cite{fazekas}. 
A spin liquid is stabilized at least at $T>0\hskip2pt$K because of the formation of resonating valence bonds between nearest-neighbor spins by the superexchange interaction. The spin liquid is the RVB spin liquid and the stabilization mechanism is the RVB mechanism.

The RVB mechanism is also effective in not only the triangular lattice but also other types of lattice in sufficiently low dimensions.
The energy scale of resonating valence bonds or local quantum spin fluctuations on nearest neighbors is $O(|J|)$.
If $T_{\rm N}\ll |J|/k_{\rm B}$, a spin state at a temperature $T$ such that $T_{\rm N}<T\ll |J|/k_{\rm B}$ is a spin liquid stabilized by the RVB mechanism, i.e., the RVB spin liquid.
Thus, it is anticipated that an electron state at the $T$ in the Hubbard model that corresponds to the Heisenberg model is an electron liquid stabilized by the RVB mechanism, i.e., the RVB electron liquid \cite{highTc,plain-vanilla}.
The RVB mechanism plays a crucial role in quenching magnetic moments, as the Kondo effect.
It is interesting to examine whether the RVB electron and spin liquids in the Hubbard and Heisenberg models are adiabatically connected to each other, as the local electron and spin liquids in the Anderson and $s$-$d$ models. 

On the other hand, the Kondo-lattice theory has been developed in previous papers \cite{Mapping-1,Mapping-2,Mapping-3,FJO-review,toyama,FJO-MottIns} in order to study strongly correlated electrons in lattice models where the onsite $U$ is crucial, such as the periodic Anderson model and the Hubbard model. 
For example, the self-energy of the Hubbard model is decomposed into the single-site and multisite self-energies; the single-site self-energy can be mapped to the local self-energy of the Anderson model; then, the multisite self-energy can be perturbatively calculated on the basis of the local self-energy of the mapped Anderson model.
On the basis of the mapping to the Anderson model,
the Kondo temperature  $T_{\rm K}$ or $k_{\rm B}T_{\rm K}$ can be defined as the energy scale of local quantum spin fluctuations in the Hubbard model. The $k_{\rm B}T_{\rm K}$ is also the energy scale of the effective Fermi energy; if $T\ll T_{\rm K}$ then electrons behave as {\it itinerant} electrons
while if $T\gg T_{\rm K}$ then electrons behave as local moments.
Thus, if $T_{\rm N}\ll T_{\rm K}$ in sufficiently low dimensions and if $T_{\rm N}<T\ll T_{\rm K}$, an electron liquid is stabilized by the Kondo effect.
The Kondo effect also plays a crucial role in suppressing magnetic instability in lattice models where the onsite $U$ is crucial, as the RVB mechanism.  
It is interesting to study by the Kondo-lattice theory how the Kondo effect and the RVB mechanism cooperate with each other to suppress magnetic instability.


One of the purposes of this paper is to study the RVB electron liquid in the Hubbard model.
The other purpose is to show that the RVB electron liquid in the Hubbard model is adiabatically connected to the RVB spin liquid in the Heisenberg model.
This paper is organized as follows:
 The Hubbard model studied in this paper is defined in \S\ref{SecModel}. 
In \S\ref{SecKL-Theory}, the Kondo-lattice theory is reformulated in order to treat properly the RVB mechanism.
In \S\ref{SecRVB},  
it is shown that the Kondo temperature is enhanced by the RVB mechanism, so that $k_{\rm B}T_{\rm K}\propto |J|$ and $k_{\rm B}T_{\rm K}= O(|J|)$ for the RVB electron liquid.
The adiabatic continuation between the RVB electron and spin liquids is examined in \S\ref{SecAdiabatic}.
Discussion is given in \S\ref{SecDiscussion}.
Conclusion is given in \S\ref{SecConclusion}.
A sum rule is proved in Appendix\hskip2pt\ref{AppSumRule}.
Two applications of the sum rule are given in Appendix\hskip2pt\ref{AppAppSumRule}.
An equality is proved in Appendix\hskip2pt\ref{AppEq}.

\section{Hubbard Model}
\label{SecModel}
The Hubbard model on a bi-partite hypercubic lattice in $D$ dimensions is studied in this paper:
\begin{equation}\label{EqHubbardH}
\mathcal{H} = 
\epsilon_d \sum_{i\sigma} n_{i\sigma}
- \frac{t}{\sqrt{D}}\sum_{ij\sigma}
\delta_{\left<ij\right>} d_{i\sigma}^\dag d_{j\sigma}^{\phantom{\dag}}
+ U \sum_{i} n_{i\uparrow} n_{i\downarrow} - \mu\mathcal{N},
\end{equation}
where $d_{i\sigma}^\dag$ and  $d_{i\sigma}^{\phantom{\dag}}$ are creation and annihilation operators of an electron at the $i$th unit cell, $n_{i\sigma}=d_{i\sigma}^\dag d_{i\sigma}^{\phantom{\dag}}$,  $\epsilon_d$ is the band center and  $\epsilon_d=0$ is assumed for simplicity, 
$-t/\sqrt{D}$ is the transfer integral between nearest neighbors,
$\delta_{\left<ij\right>}=1$ between nearest-neighbor unit cells and $\delta_{\left<ij\right>}=0$ between non-nearest-neighbor unit cells, 
$U$ is the onsite repulsion, $\mu$ is the chemical potential, and 
\begin{align}
\mathcal{N} = \sum_{i\sigma} n_{i\sigma}. 
\end{align}
The periodic boundary condition is assumed.
If $U=0$, the dispersion relation of an electron as a function of wave number ${\bm k}=(k_1,k_2, \cdots, k_D)$  is given by
\begin{align}
E({\bm k}) &=
\epsilon_d - 2t \varphi_D({\bm k}),
\quad
\varphi_D({\bm k}) =
\frac1{\sqrt{D}}\sum_{\nu=1}^{D}\cos(k_\nu a), 
\end{align} 
where $\epsilon_d=0$, as discussed above, and $a$ is the lattice constant. The density of states is given by
\begin{align}
\rho_0(\varepsilon) &=
\frac1{L}\sum_{\bm k} \delta\bigl[\varepsilon+\mu - E({\bm k})\bigr],
\end{align}
where $L$ is the number of unit cells. 
The definition of the transfer integral in Eq.\hskip2pt(\ref{EqHubbardH}) is different from that in \S\ref{SecIntroduction}; it includes the dimensional factor of $1/\sqrt{D}$, as in previous papers \cite{Metzner,Muller-H1,Muller-H2,Janis}. 
Since $(1/L)\sum_{\bm k} \varphi_D^2({\bm k})= 1/2$, 
the effective bandwidth of $E({\bm k})$ or $\rho_0(\varepsilon)$ is $O(|t|)$ for any $D$.

The electron density per unit cell is defined by
\begin{align}
n= \left<\mathcal{N}\right>\hskip-2pt/L,
\end{align}
where $\left<\cdots\right>$ stands for the statistical average.
If $\mu=\epsilon_d+U/2$, 
then $n=1$, i.e., the Hubbard model is half filled; and there exists the particle-hole symmetry in it, i.e., the Hubbard model is symmetrical.
It is assumed in this paper that $\mu$ is one such that 
$n=1$, $n=1\pm 0^+$, or $n \simeq 1$ for it, and that
$U/|t|$ is in the strong-coupling regime defined by
\begin{align}
U/|t| \gg 1.
\end{align}

The thermodynamic limit of $L\rightarrow +\infty$ is assumed.
No symmetry can be broken at nonzero temperature in one and two dimensions \cite{mermin}, and  
symmetry can be broken below a nonzero critical temperature  $T_c$ in three dimensions and higher. It is assumed that the dimensionality $D$ is sufficiently low such that $T_c$ does not exists or, if it exists, $T_c\ll T_{\rm K}$, where $T_{\rm K}$ or $k_{\rm B}T_{\rm K}$ is the energy scale discussed in \S\ref{SecIntroduction} and the definition of it is given in \S\ref{SecMapAM};
the case of $T_c\simeq T_{\rm K}$ or $T_c\gtrsim T_{\rm K}$ is out of scope in this paper.
Then, it is assumed that the temperature $T$ is within the range of
\begin{align}\label{EqTregion}
T_c+\delta T< T\lesssim T_{\rm K}, 
\end{align}
where $\delta T$ is sufficiently small such that $\delta T\ll T_{\rm K}$, but still sufficiently large; e.g., $\delta T\simeq 0.1 T_{\rm K}$, or $\delta T\simeq 0.01 T_{\rm K}$, or $\delta T\simeq 0.001 T_{\rm K}$, depending on $D$ and $\mu$, or $n$.%
\footnote{If $T_c$ does not exists, $T_c=0$ has to be assumed in Eq.\hskip2pt(\ref{EqTregion}). The $\delta T$ is introduced partly because complication arising from the logarithmic divergence of $\rho_0(\varepsilon)$ as $\varepsilon\rightarrow0$ can be avoided and partly because critical phenomena are out of scope in this paper, except for an argument in \S\ref{SecDiscussion}, in which a possible anomaly in the critical region in two dimensions is examined.} 
%
No symmetry can be broken in the temperature range defined by Eq.\hskip2pt(\ref{EqTregion}).

If no symmetry is broken, the Green function in the wave-number representation is given by
\begin{align}\label{EqGreenH}
G_\sigma(i\varepsilon_l,{\bm k}) &=
\frac1{i\varepsilon_l+ \mu 
- E({\bm k}) - \Sigma_\sigma(i\varepsilon_l,{\bm k})
- \Gamma(i\varepsilon_l)},
\end{align}
where $\varepsilon_l=(2l+1)\pi k_{\rm B}T$, with $l$ being an integer, is a fermionic energy, and $\Sigma_\sigma(i\varepsilon_l,{\bm k})$ is the self-energy. 
If $\mu=\epsilon_d+U/2$ in two dimensions, i.e., if $n=1$ and $D=2$, $\rho_0(\varepsilon)$ diverges logarithmically as $\varepsilon\rightarrow 0$. 
In order to suppress the divergence, $\Gamma(i\varepsilon_l)$ is introduced; it is due to, e.g., an explicit electron reservoir such as one considered in the previous paper \cite{toyama} or impurities considered in \S\ref{SecConductivity} of the present paper. 
For the moment, it is assumed that $\Gamma(i\varepsilon_l)$ is almost infinitesimal but still nonzero; i.e., ${\rm Im}\Gamma(\varepsilon+i0)=-0^+$ for $\varepsilon\simeq 0$, where $\Gamma(\varepsilon+i0)$ is the retarded one of $\Gamma(i\varepsilon_l)$.

The site-diagonal Green function, the density of states, and the electron density are given by
\begin{align}
&R_\sigma(i\varepsilon_l) =
\frac1{L}\sum_{\bm k} G_\sigma(i\varepsilon_l,{\bm k}),
\quad 
\rho(\varepsilon) =
- \frac1{\pi} {\rm Im} R_\sigma(\varepsilon+i0),
\quad 
n = 2 \int_{-\infty}^{+\infty} \hskip-10pt d\varepsilon 
f(\varepsilon)\rho(\varepsilon ) ,
\end{align}
respectively, where $R_\sigma(\varepsilon+i0)$ is the retarded one of $R_\sigma(i\varepsilon_l)$ and 
\begin{align}\label{EqFD-f}
f(\varepsilon) = 1/\bigl[e^{\varepsilon/(k_{\rm B}T)}+1\bigr].
\end{align}

\section{Kondo-lattice theory}
\label{SecKL-Theory}
\subsection{Mapping to the Anderson model}
\label{SecMapAM}
\subsubsection{Self-energy}
\label{SecSelf-energy}
%
The decomposition of the self-energy into the single-site and multisite ones can also be described in the wave-number representation:
$\Sigma_\sigma(i\varepsilon_l,{\bm k}) =
\Sigma_\sigma(i\varepsilon_l)
+\Delta\Sigma_\sigma(i\varepsilon_l,{\bm k})$,
where $\Sigma_\sigma(i\varepsilon_l)$ is the single-site one and $\Delta\Sigma_\sigma(i\varepsilon_l,{\bm k})$ is the multisite one.
The single-site self-energy can be mapped to the local self-energy of the Anderson model \cite{Mapping-1,Mapping-2,Mapping-3}.
The Anderson model to be mapped or the mapped Anderson model is specified by four parameters of
$\tilde{T}$, $\tilde{U}$, $\tilde{\epsilon}_d -\tilde{\mu}$, and $\tilde{\Delta}(\varepsilon)$, where  $\tilde{T}$ is the temperature of the reservoir for the Anderson model, 
$\tilde{U}$ is the onsite repulsion, which is denoted by $U$ in \S\ref{SecIntroduction}, $\tilde{\epsilon}_d$ is the level of localized electrons, $\tilde{\mu}$ is the chemical potential, and $\tilde{\Delta}(\varepsilon)$ is the hybridization energy between localized and conduction electrons.
%
The Green function, the density of states, and the density of localized electrons are given by 
\begin{align}\label{EqAnderson-G}
&\tilde{G}_\sigma(i\tilde{\varepsilon}_l) =
\left[i\tilde{\varepsilon}_l + \tilde{\mu}-\tilde{\epsilon}_d
- \tilde{\Sigma}_\sigma(i\tilde{\varepsilon}_l)
- \frac1{\pi} \int_{-\infty}^{+\infty} \hskip-10pt d \epsilon \frac{\tilde{\Delta}(\epsilon) }{i\tilde{\varepsilon}_l-\epsilon}
\right]^{-1},
\\ \label{EqDOS-A} & \hskip10pt
\tilde{\rho}(\varepsilon) =
- \frac1{\pi} {\rm Im}\hskip1pt \tilde{G}_\sigma(\varepsilon+i0),
\quad
\tilde{n} = 2 \int_{-\infty}^{+\infty} \hskip-10pt d\varepsilon 
f(\varepsilon)\tilde{\rho}(\varepsilon ) ,
\end{align}
respectively, where $\tilde{\varepsilon}_l=(2l+1)\pi k_{\rm B}\tilde{T}$, with $l$ being an integer, and $\tilde{\Sigma}_\sigma(i\tilde{\varepsilon}_l)$ is the local self-energy.
According to previous papers \cite{toyama,FJO-MottIns},
if the four parameters are given by
\begin{subequations}\label{EqMapCond}
\begin{align}\label{EqMapCond1}
& \hskip7pt 
\tilde{T}=T, \quad \tilde{U}=U,
\quad 
\tilde{\epsilon}_d -\tilde{\mu}=\epsilon_d- \mu,
\\ & \label{EqMapCond3}
\tilde{\Delta}(\varepsilon) =
{\rm Im}\left[\Sigma_\sigma(\varepsilon+i0)+1/R_{\sigma}(\varepsilon+{i}0)\right],
\end{align}
\end{subequations}
it follows that $i\tilde{\varepsilon}_l=i\varepsilon_l$ for any $l$, and 
\begin{subequations}\label{EqMap}
\begin{align}\label{EqMap1}
& \hskip55pt
\tilde{\Sigma}_\sigma(i\varepsilon_l)=\Sigma_\sigma(i\varepsilon_l),
\\ & \label{EqMap2}
\tilde{G}_\sigma(i\varepsilon_l)=R_\sigma(i\varepsilon_l),
\quad 
\tilde{\rho}(\varepsilon) =\rho(\varepsilon) ,
\quad 
\tilde{n}=n.
\end{align}
\end{subequations}
Not only the single-site $\Sigma_\sigma(i\varepsilon_l)$ but also the local $R_\sigma(i\varepsilon_l)$, $\rho(\varepsilon)$, and $n$ are mapped to the local $\tilde{\Sigma}_\sigma(i\varepsilon_l)$, $\tilde{G}_\sigma(i\varepsilon_l)$, $\tilde{\rho}(\varepsilon)$, and $\tilde{n}$, respectively, of the Anderson model.
Equation\hskip2pt(\ref{EqMapCond}) is the mapping condition.%
\footnote{In the actual mapping procedure, $\tilde{\Delta}(\varepsilon)$ has to be self-consistently determined with the single-site and multisite self-energies according to Eq.\hskip2pt(\ref{EqMapCond3}) in order for Eq.\hskip2pt(\ref{EqMap1}) to be satisfied.}
%
Since $\tilde{\Delta}(\varepsilon)$ determined by Eq.\hskip2pt(\ref{EqMapCond3}) depends on the temperature $T$ of the reservoir for the Hubbard model, the mapped Anderson model depends on the $T$, i.e., the Anderson model includes the $T$ as a {\it parameter}.
If the Hubbard model is half filled and symmetrical, the Anderson model is also half filled and symmetrical, so that $\tilde{n}=1$, $\tilde{\Delta}(\varepsilon)=\tilde{\Delta}(-\varepsilon)$, $\tilde{\rho}(\varepsilon) = \tilde{\rho}(-\varepsilon)$, ${\rm Im}\hskip1pt\tilde{G}_\sigma(\varepsilon+i0)={\rm Im}\hskip1pt\tilde{G}_\sigma(-\varepsilon+i0)$, ${\rm Re}\hskip1pt\tilde{G}_\sigma(\varepsilon+i0)=-{\rm Re}\hskip1pt\tilde{G}_\sigma(-\varepsilon+i0)$, ${\rm Re}\hskip1pt\tilde{G}_\sigma(+i0)=0$, and so on.

If the multisite self-energy is ignored in Eq.\hskip2pt(\ref{EqMapCond3}), i.e., if $\Delta\Sigma_\sigma(i\varepsilon_l,{\bm k})=0$ is assumed, the Kondo-lattice theory is reduced to the supreme single-site approximation (S$^3$A) \cite{Metzner,Muller-H1,Muller-H2,Janis}, which is rigorous for infinite dimensions or, precisely speaking, for $1/D=0$,%
\footnote{It is quite easy to show that, under S$^3$A, if Eq.\hskip2pt(\ref{EqFiniteDelta}) is satisfied then $\left[\rho(0)\right]_{T=0{\rm K}}$ is constant as a function of $U/|t|$, i.e., $\left[\rho(0)\right]_{T=0{\rm K}}=\rho_0(0)$. As studied in \S\ref{SecNature} of this paper, if the RVB mechanism is considered beyond S$^3$A and if Eq.\hskip2pt(\ref{EqFiniteDelta}) is satisfied, $\rho(0)$ decreases as $U/|t|$ increases for any finite $D$, even for $1/D=0^+$; in the half-filled case, in particular, $\rho(0)|t|\rightarrow 0$ as $U/|t|\rightarrow +\infty$ unless $1/D=0$.}
but within the constrained Hilbert subspace where no symmetry is allowed to be broken.%
\footnote{The conventional Weiss mean fields for spin density wave (SDW) or magnetism, charge density wave (CDW), and isotropic $s$-wave  or BCS superconductivity are of the zeroth order in $1/D$. Since the conventional Weiss mean fields are multisite effects, they cannot be considered in any single-site approximation.}
Either the dynamical mean-field theory \cite{georges,RevModDMFT}
or the dynamical coherent potential approximation \cite{dcpa} is the S$^3$A theory.

The mapping condition of Eq.\hskip2pt(\ref{EqMapCond3}) can also be described as 
\begin{align}\label{EqMapCondDelta}
\tilde{\Delta}(\varepsilon) &=
\frac{\pi\rho(\varepsilon)}{\bigl[{\rm Re}\hskip1ptR_\sigma(\varepsilon+i0)\bigr]^2+\bigl[\pi\rho(\varepsilon)\bigr]^2}
- \bigl|{\rm Im}\hskip1pt\Sigma_\sigma(\varepsilon+i0)\bigr|.
\end{align}
If $U/|t|=0$, $\Sigma_\sigma(\varepsilon+i0,{\bm k})=0$; and either ${\rm Re}\hskip1ptR_\sigma(\varepsilon+i0)$ or $\rho_0(\varepsilon)$ is continuous and finite at $\varepsilon=0$, even for $D=2$ and $n=1$ in the presence of $\Gamma(\varepsilon+i0)$. Then, it follows from Eq.\hskip2pt(\ref{EqMapCondDelta}) that
\begin{align}\label{EqFiniteDelta}
0 < \lim_{\varepsilon\rightarrow 0} \tilde{\Delta}(\varepsilon)<+\infty,
\end{align}
for $U=0$.
%
On the other hand, according to the previous paper \cite{toyama}, 
\begin{align}
\tilde{\Delta}(\varepsilon)\ge\bigl|{\rm Im}\hskip1pt\Gamma(\varepsilon+i0)\bigr|,
\end{align}
is satisfied for any finite $U$.
A critical $U_c$ can be defined such that Eq.\hskip2pt(\ref{EqFiniteDelta}) is satisfied for $0\le U<U_c$ while it is not satisfied for $U\ge U_c$.
In the temperature range defined by Eq.\hskip2pt(\ref{EqTregion}),
no symmetry can be broken for any finite $U$; thus,
it is anticipated that the conventional perturbative analysis in terms of $U$ is valid for any finite $U$, 
and that Eq.\hskip2pt(\ref{EqFiniteDelta}) is therefore satisfied for any finite $U$.
For the moment, it is assumed that the critical $U_c$ is infinite, even for $D=1$ and $n=1$.%
\footnote{If a complete gap opens in the self-consistent $\rho(\varepsilon)$, as proposed by Lieb and Wu \cite{lieb-wu} for the half-filled case in one dimension, Eq.\hskip2pt(\ref{EqFiniteDelta}) is not satisfied; according to the previous paper \cite{FJO-MottIns}, a necessary condition for a complete gap to open is that $\tilde{\Delta}(0)=0$ or $\tilde{\Delta}(\varepsilon)$ includes the delta function $\delta(\varepsilon)$.}
The relevance or irrelevance of this assumption is critically examined in \S\ref{SecCritialUc} and Appendix\hskip2pt\ref{AppAppSumRule}, but independently from different points of view; in particular, that in the case of $D=1$ and $n=1$ is critically  examined in \S\ref{SecDiscussion}.

\subsubsection{Polarization function in the spin channel}
The decomposition of the irreducible polarization function $\pi_s(i\omega_l,{\bm q})$ in the spin channel into the single-site and multisite ones can also be described in the wave-number representation:
$\pi_s(i\omega_l,{\bm q}) =
\pi_s(i\omega_l) + \Delta\pi_s(i\omega_l,{\bm q})$,
%
where $\pi_s(i\omega_l)$ is the single-site one, $\Delta\pi_s(i\omega_l,{\bm q})$ is the multisite one, and $\omega_l=2l\pi  k_{\rm B}T$, with $l$ being an integer, is a bosonic energy. 
The single-site $\pi_s(i\omega_l)$ is also mapped to the local $\tilde{\pi}_s(i\omega_l)$ of the Anderson model:
$\pi_s(i\omega_l)=\tilde{\pi}_s(i\omega_l)$.
The spin susceptibilities of the Anderson and Hubbard models are given by
\begin{align}
& 
\tilde{\chi}_s(i\omega_l) =
\frac{2\tilde{\pi}_s(i\omega_l)}{1-U\tilde{\pi}_s(i\omega_l)},
\quad
\chi_s(i\omega_l,{\bm q})=
\frac{2\pi_s(i\omega_l,{\bm q})}{1-U\pi_s(i\omega_l,{\bm q})},
\end{align}
respectively, 
where the conventional factor $(1/4)g^2\mu_{\rm B}^2$ is not included.

The Kondo temperature $T_{\rm K}$  can also be defined for the Hubbard model by
\begin{align}\label{EqDefTK1}
\bigl[\tilde{\chi}_s(0;T)\bigr]_{\tilde{T}\rightarrow 0\hskip2pt{\rm K}}&=1/\bigl[k_{\rm B}T_{\rm K}(T)\bigr].
\end{align}
In Eq.\hskip2pt(\ref{EqDefTK1}),
the subscript $\tilde{T}\rightarrow 0\hskip2pt{\rm K}$ means that the temperature of the reservoir for the Anderson model is zero; and 
the temperature of the reservoir for the Hubbard model, i.e., the {\it parameter} $T$ is explicitly shown because $\tilde{\chi}_s(0)$ and $T_{\rm K}$ depend on it. 
The $k_{\rm B}T_{\rm K}(T)$ or $k_{\rm B}T_{\rm K}$ defined by Eq.\hskip2pt(\ref{EqDefTK1}) is the energy scale discussed in \S\ref{SecIntroduction} and \S\ref{SecModel}.


The analysis so far is valid for any finite $U/|t|$.
The analysis in the following part is only valid for $U/|t|\gg1$.
It is obvious that if $U/|t|\gg1$ then $k_{\rm B}T_{\rm K}/U \ll 1$.
Then, it follows that
\begin{align}
&\tilde{\chi}_s(i\omega_l)=O[1/(k_{\rm B}T_{\rm K})],
\quad
\chi_s(i\omega_l,{\bm q})=O[1/(k_{\rm B}T_{\rm K})],
\end{align}
for $T\lesssim T_{\rm K}$ and $|\omega_l| \lesssim k_{\rm B}T_{\rm K}$;
and it follows that
%
\begin{align}\label{Eq-1/U}
& U\tilde{\pi}_s(i\omega_l)=1+O(k_{\rm B}T_{\rm K}/U), 
\quad
U\pi_s(i\omega_l,{\bm q})=1+O(k_{\rm B}T_{\rm K}/U),
\end{align} 
%
and, therefore, $U^2 \Delta \pi_s(i\omega_l,{\bm q})=O(1)$.
If Eq.\hskip2pt(\ref{Eq-1/U}) is used, it immediately follows that
\begin{align}
& \label{EqUPiChi} \hskip15pt
U[1-U\tilde{\pi}_s(i\omega_l)] = \bigl[2/\tilde{\chi}_s(i\omega_l)\bigr]
\bigl[1+O(k_{\rm B}T_{\rm K}/U)\bigr],
\\ & \label{EqChiKondo}
\chi_s(i\omega_l,{\bm q}) =
\frac{\tilde{\chi}_s(i\omega_l)}
{1- (1/4)I_s(i\omega_l,{\bm q})\tilde{\chi}_s(i\omega_l)}
\bigl[1 + O(k_{\rm B}T_{\rm K}/U)\bigr], 
\end{align}
for $T\lesssim T_{\rm K}$ and $|\omega_l| \lesssim k_{\rm B}T_{\rm K}$, where 
\begin{align}\label{EqExchI}
I_s(i\omega_l,{\bm q}) &= 
2U^2\Delta\pi(i\omega_l,{\bm q}).
\end{align}
%
The terms of $O(k_{\rm B}T_{\rm K}/U)$ in Eqs.\hskip2pt(\ref{EqUPiChi}) and (\ref{EqChiKondo}) are ignored in this paper.
Equation\hskip2pt(\ref{EqChiKondo}) is consistent with the physical picture for Kondo lattices that local spin fluctuations interact with each other with an intersite exchange interaction;  $I_s(i\omega_l,{\bm q})$ is none other than the intersite exchange interaction.

The N\'{e}el temperature $T_{\rm N}$ can be determined by Eq.\hskip2pt(\ref{EqChiKondo}):
\begin{subequations}\label{EqTN-Q}
\begin{align}
T_{\rm N} = \max\bigl[T_{\rm N}({\bm q})\bigr],
\end{align}
where $T_{\rm N}({\bm q})$ as a function of ${\bm q}$ is defined by
\begin{align}
\bigl[1-(1/4)I_s(0,{\bm q})\tilde{\chi}_s(0)\bigr]_{T=T_{\rm N}({\bm q})}=0.
\end{align}
\end{subequations}
In this paper, it is assumed that $D$ is sufficiently small such that $T_{\rm N}\ll T_{\rm K}$ or no $T_{\rm N}$ exists.

\subsubsection{Three-point vertex function in the spin channel}
The reducible and irreducible three-point vertex functions in the spin channel are also decomposed into the single-site and multisite ones. The single-site functions of them can also be mapped to the local vertex functions of the mapped Anderson model, respectively.
If they are denoted by
$\tilde{\Lambda}_s(i\varepsilon_l,i\varepsilon_l+i\omega_{l^\prime};i\omega_{l^\prime})$ and $\tilde{\lambda}_s(i\varepsilon_l,i\varepsilon_l+i\omega_{l^\prime};i\omega_{l^\prime})$, respectively, then
%
\begin{align}\label{EqVertexR-IR1}
\tilde{\Lambda}_s(i\varepsilon_l,i\varepsilon_l+i\omega_{l^\prime};i\omega_{l^\prime})&=
\tilde{\lambda}_s(i\varepsilon_l,i\varepsilon_l+i\omega_{l^\prime};i\omega_{l^\prime})/
\bigl[1-U\tilde{\pi}_s(i\omega_{l^\prime})\bigr].
\end{align}
If Eq.\hskip2pt(\ref{EqUPiChi}) is used, it follows that
\begin{align}\label{EqVertexR-IR2}
U\tilde{\lambda}_s(i\varepsilon_l,i\varepsilon_l+i\omega_{l^\prime};i\omega_{l^\prime}) &=
\bigl[2/\tilde{\chi}_s(i\omega_{l^\prime})\bigr]\tilde{\Lambda}_s(i\varepsilon_l,i\varepsilon_l+i\omega_{l^\prime};i\omega_{l^\prime}) .
\end{align}

\subsection{Normal Fermi liquid in the mapped Anderson model}
\label{SecNormalFL}
If Eq.\hskip2pt(\ref{EqFiniteDelta}) is satisfied, the Fermi surface exists in the conduction band of the mapped Anderson model, as discussed in Appendix\hskip2pt\ref{AppAppSumRule2}, so that definitely $T_{\rm K}>0\hskip2pt$K. If $T_{\rm K}>0\hskip2pt$K, the ground state of the Anderson model is the local normal Fermi liquid,
as discussed in \S\ref{SecIntroduction}, even if the multisite self-energy of the Hubbard model is anomalous.
%

An infinitesimal external Zeeman energy is introduced into the mapped Anderson model:
%
$\tilde{\mathcal{H}}_Z = - 
\tilde{h}(\tilde{n}_{\uparrow}-\tilde{n}_{\downarrow})$,
%
where $\tilde{h}=0^+$, and $\tilde{n}_{\sigma}$ is the number operator for localized electrons with spin $\sigma$ in the mapped Anderson model.
If $T_{\rm K}>0\hskip2pt$K, the local self-energy of the mapped Anderson model is analytic on the real axis,
%
so that it can be expanded in terms of $i\tilde{\varepsilon}_l$ \cite{yamada1,yamada2,shiba}:
\begin{align}\label{EqExpansionAM}
\tilde{\Sigma}_\sigma(i\tilde{\varepsilon}_l) &=
\tilde{\Sigma}_0
+ \bigl(1 -\tilde{\phi}_1 \bigr) i\tilde{\varepsilon}_l
+  \sigma \bigl(1-\tilde{\phi}_s\bigr) \tilde{h}
- \tilde{\phi}_{2} (i\tilde{\varepsilon}_l)^2/\bigl[\pi\tilde{\Delta}(0)\bigr]
\nonumber \\ & \quad
- i (\tilde{\varepsilon}_l/|\tilde{\varepsilon}_l|)\bigl[\tilde{\phi}_{21}(i\tilde{\varepsilon}_l)^2
+\tilde{\phi}_{22} (k_{\rm B}\tilde{T})^2\bigr]/\bigl[\pi\tilde{\Delta}(0)\bigr]
+\cdots,
\end{align}
where 
$\tilde{\phi}_1\ge 1$, $\tilde{\phi}_s\ge1$, $\tilde{\phi}_{2}\gtrless 0$, $\tilde{\phi}_{21}\ge 0$, and $\tilde{\phi}_{22}\ge 0$, in general; and $\tilde{\phi}_{2}= 0$ in the symmetrical model. If $\tilde{U}/[\pi\tilde{\Delta}(0)]\gg 1$, then $\tilde{\phi}_1 \gg 1$, $\tilde{\phi}_{21} \gg 1$, and $\tilde{\phi}_{22} \gg 1$; and 
\begin{align}\label{Eq3W}
\tilde{W}_s = \tilde{\phi}_s/\tilde{\phi}_1=O(1),  \quad
\tilde{W}_{21} =\tilde{\phi}_{21}/\tilde{\phi}_1^2=O(1), \quad
\tilde{W}_{22} =\tilde{\phi}_{22}/\tilde{\phi}_1^2=O(1). 
\end{align}
The $\tilde{W}_s$ is none other than the Wilson ratio.
If $\tilde{\Delta}(\varepsilon)$ is constant as a function of $\varepsilon$, $\tilde{W}_s=2$ in the $s$-$d$ model or the $s$-$d$ limit of the Anderson model \cite{wilsonKG, yamada1,yamada2}. It is anticipated that if $U/|t|\gg1 $ and $n\simeq 1$ then $\tilde{W}_s\simeq 2$ for the mapped Anderson model, whose $\tilde{\Delta}(\varepsilon)$ depends on $\varepsilon$, in general.

Since the self-energy $\tilde{\Sigma}_\sigma(i\tilde{\varepsilon}_l)$ is analytic on the real axis,
the Fermi-liquid relation \cite{Luttinger1,Luttinger2,yamada1,yamada2} has to be satisfied in a self-consistent solution, even if the multi-site $\Delta\Sigma_\sigma(\varepsilon+i0,{\bm k})$ is anomalous. 
Here, $\tilde{T}$ is treated as being independent of $T$; the parameter $T$ is explicitly shown.
If $\tilde{T}\rightarrow 0\hskip2pt$K is assumed for a given $T$, the susceptibility of the Anderson model is given by
\begin{align}\label{EqSusAM}
\bigl[\tilde{\chi}_s(0;T)\bigr]_{\tilde{T}\rightarrow0\hskip2pt{\rm K}} = 1/\left[k_{\rm B}T_{\rm K}(T)\right]
=\bigl[2 \tilde{\phi}_s(T)\tilde{\rho}(0;T)\bigr]_{\tilde{T}\rightarrow0\hskip2pt{\rm K}},
\end{align}
where Eq.\hskip2pt(\ref{EqDefTK1}) is used.
If $\tilde{T}$ is sufficiently low such that $\tilde{T}\ll T_{\rm K}$, then it follows that
\begin{align}\label{EqSusAM1}
\tilde{\chi}_s(0;T) &= 
\bigl[1-O\bigl(\delta_{\tilde{T}}^2\bigr)\bigr]\big/\bigl[k_{\rm B}T_{\rm K}(T)\bigr]
= 2 \tilde{\phi}_s(T)\tilde{\rho}(0;T) 
\bigl[1-O\bigl(\delta_{\tilde{T}}^2\bigr)\bigr],
\end{align}
where $\delta_{\tilde{T}}= \tilde{T}/T_{\rm K}(T)$; therefore, it follows that
\begin{align}\label{EqRhoNum1}
\tilde{\rho}(0;T) \simeq 1/\bigl[2\tilde{\phi}_s(T)k_{\rm B}T_{\rm K}(T)\bigr] \propto 1/\bigl[\tilde{\phi}_1(T) k_{\rm B}T_{\rm K}(T)\bigr].
\end{align}
On the other hand, if $\tilde{n}=1$ and $\tilde{T}\rightarrow 0\hskip2pt$K, then ${\rm Re}\tilde{G}_\sigma(+i0;T)=0$ and ${\rm Im}\tilde{\Sigma}_\sigma(+i0;T)\rightarrow 0$; thus, it follows that according to Eqs.\hskip2pt(\ref{EqMap}) and (\ref{EqMapCondDelta}) that
\begin{align}\label{EqFLR2}
\bigl[\pi\tilde{\rho}(0;T)\tilde{\Delta}(0;T)\bigr]_{\tilde{T}\rightarrow0\hskip2pt{\rm K}} =1.
\end{align}
Therefore, if $\tilde{T}\ll T_{\rm K}(T)$ and $\tilde{n}\simeq 1$ then $\pi\tilde{\rho}(0;T)\tilde{\Delta}(0;T)\simeq 1$, so that
\begin{align}\label{EqD-Pi}
\pi\tilde{\Delta}(0;T) \simeq 2\tilde{\phi}_s(T)k_{\rm B}T_{\rm K}(T) \propto \tilde{\phi}_1(T) k_{\rm B}T_{\rm K}(T).
\end{align}
The bosonic energy for $\tilde{T}$ is denoted by $\tilde{\omega}_{l}=2\pi l k_{\rm B}\tilde{T}$, with $l$ being an integer.
According to the Ward relation \cite{ward}, it follows that
%
\begin{align}\label{EqWard1}
\tilde{\Lambda}_s(i\tilde{\varepsilon}_l,i\tilde{\varepsilon}_l+i\tilde{\omega}_{l^\prime};i\tilde{\omega}_{l^\prime};T)&=
1- \lim_{h\rightarrow 0}\frac1{2}
\bigl[d/(d \tilde{h})\bigr] \bigl[\tilde{\Sigma}_\uparrow(i\tilde{\varepsilon}_l;T)-\tilde{\Sigma}_\downarrow(i\tilde{\varepsilon}_l;T)\bigr],
\end{align}
where $\tilde{\omega}_{l^\prime}=0$.
According to Eqs.\hskip2pt(\ref{EqExpansionAM}) and (\ref{EqWard1}),
\begin{align}\label{EqWard2}
\tilde{\Lambda}_s(i\tilde{\varepsilon}_l,i\tilde{\varepsilon}_l+i\tilde{\omega}_{l^\prime};i\tilde{\omega}_{l^\prime};T)
&= \tilde{\phi}_s(T) ,
\end{align}
where $\tilde{\omega}_{l^\prime}=0$. 
According to Eqs.\hskip2pt(\ref{EqVertexR-IR2}) and (\ref{EqWard2}), 
\begin{align}\label{EqUlambda}
U\tilde{\lambda}_s(i\tilde{\varepsilon}_{l},i\tilde{\varepsilon}_{l}+i\tilde{\omega}_{l^\prime};i\tilde{\omega}_{l^\prime};T)=2\tilde{\phi}_s(T)/\tilde{\chi}_s(i\tilde{\omega}_{l^\prime};T),
\end{align}
where $\tilde{\omega}_{l^\prime}=0$.
In this paper, Eq.\hskip2pt(\ref{EqUlambda}) is used for $|\tilde{\varepsilon}_{l}| \lesssim k_{\rm B}T_{\rm K}(T)$ and $|\tilde{\omega}_{l^\prime}|\lesssim k_{\rm B}T_{\rm K}(T)$.

When the above relations are used in a self-consistent solution for the Hubbard model, $\tilde{T}=T$ has to be assumed.
Every single-site property depends on the {\it parameter} $T$.
If the {\it parameter} $T$ is much lower than $T_{\rm K}$, the {\it parameter}-$T$ dependence is so small that it can be ignored, except in the case of $D=2$ and $n\simeq 1$, as discussed in \S\ref{SecDiscussion}. 

If Eq.\hskip2pt(\ref{EqExpansionAM}) is used, the Green function of the Hubbard model is given by
\begin{subequations}\label{EqGreenKL}
\begin{align}\label{EqGreenKL1}
& \hskip65pt
G_\sigma (i\varepsilon_l,{\bm k}) = (1/\tilde{\phi}_1)g_\sigma (i\varepsilon_l,{\bm k}),
\\  \label{EqGreenKL2}
& g_\sigma (i\varepsilon_l,{\bm k}) =
\frac1{i\varepsilon_l+\mu^*- \left[E({\bm k}) 
+ \Delta\Sigma_\sigma(i\varepsilon_l,{\bm k})\right]/\tilde{\phi}_1
- \tilde{\gamma}_{\rm K}(i\varepsilon_l) } ,
%
%
\\ & \label{EqGreenKL3} \hskip90pt
\mu^* = \bigl[\mu-\tilde{\Sigma}_0)\bigr]/\tilde{\phi}_1,
\\ & \label{EqGreenKL4} \hskip3pt
\tilde{\gamma}_{\rm K}(i\varepsilon_l) = 
- i(\varepsilon_l/|\varepsilon_l|)
\bigl[\tilde{W}_{21}(i\varepsilon_l)^2 
+\tilde{W}_{22} (k_{\rm B}T)^2\bigr]
\tilde{\phi}_1/\bigl[\pi\tilde{\Delta}(0)\bigr].
\end{align}
\end{subequations}
In Eq.\hskip2pt(\ref{EqGreenKL2}), $\Gamma(i\varepsilon_l)$, $\tilde{h}$, and $- \tilde{\phi}_2 (i\varepsilon_l)^2/\bigl[\pi\tilde{\Delta}(0)\bigr]$ are ignored. 
The Green function given by Eq.\hskip2pt(\ref{EqGreenKL}) can describe only electrons in the vicinity of the chemical potential but it cannot describe electrons in the upper and lower Hubbard bands; Eq.\hskip2pt(\ref{EqGreenKL2}) is accurate for $|\varepsilon_l|\ll k_{\rm B}T_{\rm K}$ and $T\ll T_{\rm K}$ and is approximate for $|\varepsilon_l|\lesssim k_{\rm B}T_{\rm K}$ or $T\lesssim T_{\rm K}$.
If $0<1/\tilde{\phi}_1\ll 1$ and $0<k_{\rm B}T_{\rm K}\ll |t|$, a narrow band appears in the vicinity of the chemical potential.
The narrow band is none other than the Gutzwiller band \cite{Gutzwiller1,Gutzwiller2,Gutzwiller3}.
The spectral weight of the Gutzwiller band is $1/\tilde{\phi}_1$.

\subsection{Superexchange Interaction}
\label{SecJs}
The intersite exchange Interaction $I(i\omega_l,{\bm q})$ can be decomposed into three terms: 
\begin{align}\label{EqThreeJ}
I_s(0,{\bm q})= J_s(0,{\bm q})+ J_Q(0,{\bm q}) +\Lambda(0,{\bm q}), 
\end{align}
where $J_s(0,{\bm q})$ is the superexchange interaction, and $J_Q(0,{\bm q};T)$ is an exchange interaction that 
arises from the virtual exchange of a pair excitation of an electron and a hole within the Gutzwiller band \cite{satoh,miyai,FJO-review},%
\footnote{According to a previous paper \cite{satoh}, $J_Q(0,{\bm q})$ has a novel property such that its strength is proportional to $k_{\rm B}T_{\rm K}$, i.e., the bandwidth of the Gutzwiller band.}
%
and $\Lambda(i\omega_l,{\bm q})$ is the sum of all the remaining terms of $I_s(i\omega_l,{\bm q})$, such as the mode-mode coupling term, terms due to quantum and thermal critical fluctuations, and so on.
Only the superexchange interaction $J_s(i\omega_l,{\bm q})$ is  considered in this subsection.
%


According to Hubbard's theory \cite{hubbard1,hubbard3}, the band splits into the upper and lower Hubbard bands.
Since Hubbard's theory is under the single-site approximation, this result on the band splitting can be approximately used for high-energy local properties of the mapped Anderson model. 
The Green function and the self-energy of the mapped Anderson model are approximately given by
%
%
\begin{subequations}\label{EqLocalG-H}
\begin{align}\label{EqLocalG-H1}
\tilde{G}_\sigma(i\varepsilon_l) &=
\frac1{i\varepsilon_l+\mu-\epsilon_d + \sigma \tilde{h}
- \tilde{\Sigma}_\sigma(i\varepsilon_l) }
\\ \label{EqLocalG-H2} &
= \frac{1-\left<\tilde{n}_{-\sigma}\right>}{i\varepsilon_l+\mu-\epsilon_d}
+ \frac{\left<\tilde{n}_{-\sigma}\right>}{i\varepsilon_l+\mu-\epsilon_d-U},
\end{align}
\end{subequations}
for $|\varepsilon_l| \gg k_{\rm B}T_{\rm K}$, where $\left<\tilde{n}_\sigma\right>$ is the number of localized electrons with spin $\sigma$ in the presence of the infinitesimal Zeeman energy $\tilde{h}$; in Eq.\hskip2pt(\ref{EqLocalG-H2}), $\tilde{h}$'s in the denominators are ignored because they are not crucial.
If the rigorous $\left<\tilde{n}_{\uparrow}\right>$ and $\left<\tilde{n}_{\downarrow}\right>$ are used in Eq.\hskip2pt(\ref{EqLocalG-H}), i.e., if
\begin{align}
\tilde{\chi}_s(0) = \lim_{\tilde{h}\rightarrow 0}
(d/d\tilde{h})
\bigl(\left<\tilde{n}_{\uparrow}\right>-\left<\tilde{n}_{\downarrow}\right>\bigr),
\end{align}
is used, it follows according to Eqs.\hskip2pt(\ref{EqVertexR-IR2}), (\ref{EqWard1}), and (\ref{EqLocalG-H}) that
\begin{align}\label{EqVertexHigh}
U\tilde{\lambda}_s(i\varepsilon_l,i\varepsilon_l+i\omega_{l^\prime};i\omega_{l^\prime}) &= 
- \frac1{\tilde{G}_\sigma^2(i\varepsilon_l)}\left(
\frac1{i\varepsilon_l+\mu-\epsilon_d}
+\frac1{i\varepsilon_l+\mu-\epsilon_d-U}\right),
\end{align}
for $\omega_{l^\prime}=0$ and $|\varepsilon_l| \gg k_{\rm B}T_{\rm K}$.

The superexchange interaction arises from the virtual exchange of an electron in the upper Hubbard band and a hole in the lower Hubbard band, as discussed in \S\ref{SecIntroduction}, and 
is a second-order effect in $-t/\sqrt{D}$.
According to Eq.\hskip2pt(\ref{EqExchI}), therefore, the superexchange interaction is given by
\begin{subequations}
\begin{align}\label{EsSiteJs}
& \hskip110pt
J_s(i\omega_l,{\bm q}) =
\frac1{L}\sum_{\bm q} e^{i{\bm q}\cdot({\bm R}_i-{\bm R}_j)}J_{ij}(i\omega_l),
%
\\ &
J_{ij}(i\omega_l) =
\delta_{\left<ij\right>} 2 k_{\rm B}T \sum_{l^\prime}
\left(-t/\sqrt{D}\right)^2 U^2\tilde{\lambda}_s^2(i\varepsilon_{l^\prime},i\varepsilon_{l^\prime}+i\omega_l;i\omega_{l})
R_\sigma^2(i\varepsilon_{l^\prime})R_\sigma^2(i\varepsilon_{l^\prime}+i\omega_l),
\end{align}
\end{subequations}
where $R_\sigma(i\varepsilon_l)=\tilde{G}_\sigma(i\varepsilon_l)$.
If Eq.\hskip2pt(\ref{EqVertexHigh}) is used, the static part of $J_{ij}(i\omega_l)$ is given by 
\begin{align}\label{EqJ_ij}
J_{ij}(0) =\delta_{\left<ij\right>}J/D,
\quad
J=-4t^2/U.
\end{align}
This agrees with the one derived by the conventional theory \cite{Js-mech-pert}.

Since 
$J_{ij}(\omega+i0)\rightarrow 0$ as $\omega\rightarrow+\infty$, and $J_{ij}(\omega+i0)$ is analytical in the upper-half complex plane, $J_{ij}(i\omega_l)$ can be described, in general, as
\begin{align}
J_{ij}(i\omega_l) = \delta_{\left<ij\right>}\frac{J}{D}\hskip-3pt
\int_{0}^{+\infty} \hskip-10pt dx X_J(x)\left(\frac1{i\omega_l+x}-\frac1{i\omega_l-x}\right),
\quad 
\int_{0}^{+\infty} \hskip-10pt dx \frac{X_J(x)}{x} =\frac1{2}.
\end{align}
Since $X_J(x)$ has a peak around $x=U$, it is assumed in this paper that
%
$X_J(x) = (1/2)U\delta(x-U)$.
%
Then, it follows that
%
\begin{align}\label{EqJs}
J_s(i\omega_l,{\bm q}) &=
\frac{2J}{\sqrt{D}} \varphi_D({\bm q})\hskip1pt
\frac{U}{2}\left(\frac1{i\omega_l+U}-\frac1{i\omega_l-U}\right).
\end{align}
In the static limit of $|\omega_l|/U\rightarrow 0$, Eq.\hskip2pt(\ref{EqJs}) is reduced to
\begin{align}\label{EqJs2}
J_s(i\omega_l,{\bm q}) =
\frac{2J}{\sqrt{D}}\varphi_D({\bm q})
= \frac{2J}{D}\sum_{\nu=1}^{D}\cos(q_\nu a).
\end{align}
The $J_s(i\omega_l,{\bm q})$ is of higher order in $1/D$ for almost all the ${\bm q}$'s, although it is of the zeroth order in $1/D$ for particular ${\bm q}$'s such as ${\bm q}=(0,0, \cdots, 0)$, ${\bm q}=(\pi/a)(\pm1, \pm1, \cdots, \pm1)$, and so on.

\subsection{Perturbative scheme of the Kondo-lattice theory}
Electrons in the vicinity of $\mu$ interact with each other by the mutual interaction mediated by spin fluctuations.
The mutual interaction of the first order in the spin-fluctuation mode is given by
\begin{align}\label{EqI*1}
U^2\tilde{\lambda}_s(i\varepsilon_{l^\prime},i\varepsilon_{l^\prime}+i\omega_{l};i\omega_{l})
\tilde{\lambda}_s(i\varepsilon_{l^{\prime\prime}},i\varepsilon_{l^{\prime\prime}}-i\omega_{l};-i\omega_{l})
\left[\chi_s(i\omega_{l},{\bm q})-\tilde{\chi}_s(i\omega_{l})\right],
\end{align}
where $i\varepsilon_{l^\prime}$ and $i\varepsilon_{l^{\prime\prime}}$ are the energies of incoming electrons, and $i\varepsilon_{l^{\prime}}+i\omega_{l}$ and $i\varepsilon_{l^{\prime\prime}}-i\omega_{l}$ are the energies of outgoing electrons. 
Since the single-site part is considered in the mapping to the Anderson model, it is subtracted in Eq.\hskip2pt(\ref{EqI*1}) in order to avoid double counting. 
It follows that
\begin{subequations}\label{EqChiI*}
\begin{align}
& \label{EqChiI*1}
\chi_s(i\omega_l,{\bm q})-\tilde{\chi}_s(i\omega_l) =
(1/4)\tilde{\chi}^2(i\omega_l) I_s^*(i\omega_l,{\bm q}),
\\ & \label{EqChiI*2} \hskip10pt
I_s^*(i\omega_l,{\bm q})=
\frac{I_s(i\omega_l,{\bm q})}{1- (1/4)I_s(i\omega_l,{\bm q})\tilde{\chi}_s(i\omega_l)}.
\end{align}
\end{subequations}
%
%
If Eqs.\hskip2pt(\ref{EqUlambda}) and (\ref{EqChiI*}) are used for $|\omega_l|\lesssim k_{\rm B}T_{\rm K}$, Eq.\hskip2pt(\ref{EqI*1}) is simply described as 
$\tilde{\phi}_s^2 I_s^*(i\omega_l,{\bm q})$.

If once the mapped Anderson model is solved and the single-site $\tilde{\Sigma}_\sigma(i\varepsilon_l)$ and $\tilde{\chi}_s(i\omega_l)$ are given, the multisite $\Delta\Sigma_\sigma(i\varepsilon_l,{\bm k})$ and $\Delta\pi(i\omega_l,{\bm q})$ can be perturbatively calculated in terms of the intersite $I_s(i\omega_l,{\bm q})$, although the mapped Anderson model has to be self-consistently solved with the multisite terms.
Since the single-site terms are given by the local terms of the mapped Anderson model, only multisite terms have to be considered in this perturbative scheme in order to avoid double counting.
In this perturbative scheme, thus, the intersite $I_s(i\omega_l,{\bm q})$ has to be treated as the {\it bare} intersite exchange interaction, and 
the single-site $\tilde{\phi}_s$ has to be treated as the {\it bare} vertex function in the spin channel;
and $I_s^*(i\omega_l,{\bm q})$ is the renormalized intersite exchange interaction, which is enhanced or screened by intersite spin fluctuations depending on ${\bm q}$.

It is straightforward to show that $I_s(i\omega_l,{\bm q})$ is of higher order in $1/D$ for almost all the ${\bm q}$'s except for particular ${\bm q}$'s, as $J_s(i\omega_l,{\bm q})$; e.g., 
the $I_s(0,{\bm Q})$, where ${\bm Q}$ is the ordering wave number determined by Eq.\hskip2pt(\ref{EqTN-Q}), is of the zeroth order in $1/D$ and it corresponds to the conventional Weiss mean field.
The Kondo-lattice theory, which is a perturbative theory in terms of $I_s(i\omega_l,{\bm q})$, is none other than $1/D$ expansion theory.

\section{Resonating-Valence-Bond (RVB) Electron Liquid}
\label{SecRVB}
\subsection{Fock-type exchange effect of the superexchange interaction}
\label{SecRVB-Single}
There are two types of self-energy of the first order in $J_s(i\omega_l,{\bm q})$. One is the Hartree-type self-energy:
\begin{align}\label{EqHartree}
&\Sigma_\sigma^{(J_s)}(i\varepsilon_l) =
k_{\rm B}T \sum_{l^\prime}
e^{i\varepsilon_{l^\prime} 0^+} \tilde{\phi}_s \frac{1}{4}J_s(0,0)
R_{-\sigma}({i}\varepsilon_{l^\prime}).
\end{align}
The other is the Fock-type self-energy:
\begin{align}\label{EqFock}
\Delta\Sigma_\sigma^{\rm (RVB)}(i\varepsilon_l,{\bm k}) =
\frac{k_{\rm B}T}{L}
\sum_{l^\prime{\bm p}\sigma^\prime}
\tilde{\phi}_s^2 \frac{1}{4}J_s(i\varepsilon_l-i\varepsilon_{l^\prime},{\bm k}-{\bm p})
\bigl({\bm \sigma}^{\sigma\sigma^\prime}
\hskip-4pt\cdot\hskip-1pt{\bm \sigma}^{\sigma^\prime\hskip-1pt\sigma}\bigr)
G_{\sigma^\prime}({i}\varepsilon_{l^\prime},{\bm p}),
\end{align}
where ${\bm \sigma}=(\sigma_x,\sigma_y,\sigma_z)$ is the Pauli matrix.
The Hartree-type self-energy is included in the conventional Hartree term, which is given by 
\begin{align}
\Sigma_\sigma^{({\rm H})}(i\varepsilon_l)=k_{\rm B}T \sum_{l^\prime} e^{i\varepsilon_{l^\prime} 0^+}U R_{-\sigma}(i\varepsilon_{l^\prime}),
\end{align}
as a part of it.
Since the conventional Hartree term is one of the terms for the single-site self-energy, which is considered in the mapping to the Anderson model, the Hartree-type self-energy has not to be considered in order to avoid double counting.
What is considered by the Fock-type self-energy is the Fock-type exchange effect of the superexchange interaction, which is none other than the RVB mechanism \cite{plain-vanilla}; thus, the Fock-type self-energy is called the RVB self-energy.

If Eqs.~(\ref{EqGreenKL}) and (\ref{EqJs}), and the equality of  
\begin{align}\label{EqJsKP}
2\varphi_D({\bm k}-{\bm p}) &=
\frac1{\sqrt{D}}\sum_{\nu=1}^{D}\bigl[
\cos(k_\nu a)\cos(p_\nu a)
+ \sin(k_\nu a)\sin(p_\nu a)\bigr],
\end{align}
are used, the RVB self-energy is calculated to be
\begin{align}\label{EqFock1}
\Delta\Sigma_\sigma^{\rm (RVB)}(i\varepsilon_l,{\bm k}) &=
(3/4)\tilde{\phi}_1 \tilde{W}_s^2 (J/D)\Xi_D(i\varepsilon_l)\varphi_D({\bm k}),
\end{align}
where $\tilde{W}_s=\tilde{\phi}_s/\tilde{\phi}_1$ is the Wilson ratio, and 
\begin{align}
\Xi_D(i\varepsilon_l) &=
\frac1{L}\sum_{{\bm p}}\varphi_D({\bm p})
 \Biggl\{\frac{U}{2}\Bigl[
n(U)g_\sigma(i\varepsilon_l+U,{\bm p}) - n(-U)g_\sigma(i\varepsilon_l-U,{\bm p}) \Bigr]
\nonumber \\ & \qquad
- \frac1{\pi}\int_{-\infty}^{+\infty} \hskip-10pt d\epsilon f(\epsilon)
\hskip1pt \frac{U}{2} \hskip-2pt
\left(\frac1{i\varepsilon_l-\epsilon + U }
-\frac1{i\varepsilon_l -\epsilon - U}\right)\hskip-1pt
{\rm Im} g_\sigma(\epsilon+i0,{\bm p}) \Biggr\},
\end{align}
where $g_\sigma(\epsilon+i0,{\bm p})$ is given by Eq.\hskip2pt(\ref{EqGreenKL2}), $f(\epsilon)$ is defined by Eq.\hskip2pt(\ref{EqFD-f}), and
\begin{align}
n(\epsilon)=1/\bigl[e^{\epsilon/(k_{\rm B}T)} - 1\bigr].
\end{align}
It is easy to show that 
%
\begin{align}\label{EqRVB-Vanish}
\lim_{|\varepsilon_l|/U\rightarrow +\infty}\Xi_D(i\varepsilon_l)=0, 
\quad
\lim_{|\varepsilon_l|/U\rightarrow +\infty}\Delta\Sigma_\sigma^{\rm (RVB)}(i\varepsilon_l,{\bm k})=0.
\end{align}
In the large limit of $U/|\varepsilon_l|$,  
$\Xi_D(i\varepsilon_l)$ is simply given by
\begin{align}\label{EqXiD}
\Xi_D & =
\frac1{L}\sum_{{\bm p}}\varphi_D({\bm p})
\int_{-\infty}^{+\infty}\hskip-10pt d\epsilon 
f(\epsilon) \left(-\frac1{\pi}\right) 
{\rm Im}\hskip1pt g_\sigma(\epsilon+i0,{\bm p}).
\end{align}
%
If $U/|t|\gg 1$ and $|\varepsilon_l|/U \ll 1$, Eq.\hskip2pt(\ref{EqXiD}) can be used for $\Xi_D(i\varepsilon_l)$ with a sufficient accuracy. Then,
%
\begin{align}\label{EqRVB-Self}
\Delta\Sigma_\sigma^{\rm (RVB)}(i\varepsilon_l,{\bm k}) & =
\tilde{\phi}_1 (3/4) \tilde{W}_s^2  (J/D) (t/|t|)|\Xi_D|\varphi_D({\bm k}).
\end{align}
%
Even if Eq.\hskip1pt(\ref{EqJs2}) is used instead of Eq.\hskip2pt(\ref{EqJs}),
Eq.\hskip2pt(\ref{EqRVB-Self}) can also be derived.
The RVB self-energy is of higher order in $1/D$, as anticipated.

Since $\Delta\Sigma_\sigma^{\rm (RVB)}(i\varepsilon_l,{\bm k})$ given by Eq.\hskip2pt(\ref{EqRVB-Self}) does not depend on $i\varepsilon_l$, 
$g_{\sigma}({i}\varepsilon_l,{\bm k})$ is simply given by
\begin{subequations}\label{EqGr-KL}
\begin{align}
& \label{EqGr-KL1} \hskip90pt 
g_{\sigma} (i\varepsilon_l,{\bm k}) =
\frac1{i\varepsilon_l+\mu^*-\xi({\bm k})
-\tilde{\gamma}_{\rm K}(i\varepsilon_l)},
\\ \label{EqGr-KL2} & \hskip0pt
\xi({\bm k}) = -2t^* \varphi_D({\bm k}),
%
\quad
2t^* = 2(t/\tilde{\phi}_1) + (t/|t|) c_J (|J|/D) ,
\quad
c_J = (3/4) \tilde{W}_s^2 |\Xi_D|,
\end{align}
\end{subequations}
where $\mu^*$ is defined by Eq.\hskip2pt(\ref{EqGreenKL3}) and $|J|=4t^2/U$.
The density of states is given by 
\begin{align}\label{EqFLRho}
\rho(\varepsilon) &= 
\tilde{\rho}(\varepsilon) =
- \frac1{\pi  L} \sum_{\bm k} 
{\rm Im} \left[\bigl(1/\tilde{\phi}_1\bigr) g_{\sigma}(\varepsilon+{i}0,{\bm k}) \right].
\end{align}
Either Eq.\hskip2pt(\ref{EqGr-KL}) or (\ref{EqFLRho}) can describe only the Gutzwiller band but cannot describe the upper and lower Hubbard bands.

Since $c_J$ or $|\Xi_D|$, $\mu^*$, $t^*$, and $\tilde{\phi}_1$ depend on each other, they have to be self-consistently calculated with each other as a funciton of $T$ and $\mu$, or $T$ and $n(\mu)$.
If no symmetry is broken even at $T=0\hskip2pt$K, 
the Fermi-surface sum rule \cite{Luttinger1,Luttinger2} can be used to determine $\mu^*$ for $T=0\hskip2pt$K:
\begin{align}\label{EqFS-SumRule}
n(\mu) = \frac{2}{L}\sum_{\bm k}
\int_{-\infty}^{+\infty} \hskip-10pt d\epsilon f(\epsilon)\delta\bigl[\epsilon+\mu^*-\xi({\bm k})\bigr].
\end{align}
If $T=0\hskip2pt$K is assumed in Eq.\hskip2pt(\ref{EqFS-SumRule}), Eq.\hskip2pt(\ref{EqFS-SumRule}) is none other than the Fermi-surface sum rule; e.g., certainly $\mu^*=0$ for $n=1$, which is required to be satisfied because of the particle-hole symmetry.
If $T=0\hskip2pt$K and $\mu^*=0$ are assumed, it is easy to calculate $\Xi_D$ defined by Eq.\hskip2pt(\ref{EqXiD}); e.g., $|\Xi_{1}|=1/\pi=0.31831\cdots$, $|\Xi_{2}|=2\sqrt{2}/\pi^2=0.28658\cdots$, $\cdots$, and $|\Xi_{\infty}|=1/(2\sqrt{\pi})=0.283095\cdots$; 
i.e., $|\Xi_{D}|\simeq 1/3$ for any $D$. 
If $|\Xi_{D}|= 1/3$ and $\tilde{W}_s=2$ are assumed, then $c_J =1$.
%
Since $|t^*| \propto k_{\rm B}T_{\rm K}$ and $|t^*| =O( k_{\rm B}T_{\rm K})$, as shown in Eq.\hskip2pt(\ref{EqTKNum2}) in \S\ref{SecNature}, 
Eq.\hskip2pt(\ref{EqFS-SumRule}) can be used, approximately but with a sufficient accuracy, for $T$ in the range defined by Eq.\hskip2pt(\ref{EqTregion}).
If $n\simeq 1$ and $T$ is in the range defined by Eq.\hskip2pt(\ref{EqTregion}), then $|\mu^*|\ll |t^*|$, so that $|\Xi_{D}|\simeq 1/3$ and $c_J\simeq 1$ or $c_J=O(1)$.

\subsection{Nature of the RVB electron liquid}
\label{SecNature}
%
The single-site self-energy has to be self-consistently calculated or determined with the RVB self-energy, although the self-cosnsistency is not completed in \S\ref{SecRVB-Single}.
In this subsection,
the nature of the self-consistent solution is studied under the assumption that the self-consistentcy is completed.

If the self-consistent $\tilde{\phi}_1$ is determined for a $U$ such that $U/|t|\gg 1$, then, according to Eq.\hskip2pt(\ref{EqGr-KL2}),
\begin{align}\label{Eq-t*-t}
|t^*|= |t|\bigl[(1/\tilde{\phi}_1) + 2c_J |t|/(DU)\bigr],
\end{align}
where $c_J=O(1)$.
If $U/|t|\gg 1$ and if $n=1$ or $n\simeq 1$, then $1/\tilde{\phi}_1\ll 1$. Thus, it follows that
\begin{align}\label{EqMidBandW1}
(1/\tilde{\phi}_1) + 2c_J |t|/(DU)\ll 1, 
\quad 
|t^*| \ll  |t|.
\end{align}
Since $|t^*| \ll  |t|$, the Gutzwiller band is certainly formed.
Its bandwidth is $O(|t^*|)$.

If $U/|t|\gg 1$ and $1/\tilde{\phi}_1$ is one such that it satisfies 
\begin{align}\label{EqRVB-Q1}
2c_J |t|/(DU) \ll 1/\tilde{\phi}_1 \ll 1,
\end{align}
the RVB self-energy is not the main term of Eq.\hskip2pt(\ref{Eq-t*-t}).
The bandwidth of the Gutzwiller band, which is $O(|t^*|)$, decreases as $1/\tilde{\phi}_1$ decreases. 
If $1/\tilde{\phi}_1$ is nonzero in the limit of $U/|t|\rightarrow +\infty$, as shown in Eq.\hskip2pt(\ref{EqGutzwiller}) in \S\ref{SecCritialUc}, the spectral weight of the Guztwiller band, $1/\tilde{\phi}_1$, is also nonzero even in the limit of $U/|t|\rightarrow +\infty$.

On the other hand, if $U/|t|\gg 1$ and $1/\tilde{\phi}_1$ is so small that it satisfies 
\begin{align}\label{EqRVB-Q2}
1/\tilde{\phi}_1 \lesssim 2c_J |t|/(DU)\ll 1,
\end{align}
the RVB self-energy is the main term of Eq.\hskip2pt(\ref{Eq-t*-t}).
Since the electron liquid in the Gutzwiller band is mainly stabilized by the RVB mechanism, it is none other than the RVB electron liquid.
The bandwidth of the Gutzwiller band or
the energy scale of the RVB electron liquid is almost independent of $1/\tilde{\phi}_1$ and is $O\bigl[t^2/(DU)\bigr]$.
However, the spectral weight of the Gutzwiller band, $1/\tilde{\phi}_1$, is quite small or almost vanishing. 
Thus, the RVB electron liquid is almost a spin liquid or a quasi spin liquid.

In the temperature range defined by Eq.\hskip2pt(\ref{EqTregion}),
the Fermi-liquid relations discussed in \S\ref{SecNormalFL} are  satisfied, approximately but with a sufficient accuracy for the analysis of this paper.
If Eq.\hskip2pt(\ref{EqRVB-Q1}) or (\ref{EqRVB-Q2}) is satisfied, it follows according to Eqs.\hskip2pt(\ref{EqGr-KL}) and (\ref{EqFLRho}) that
\begin{align}\label{EqRhoNum2}
\rho(0) \propto 1/(\tilde{\phi}_1|t^*|),
\quad
\rho(0) =O\bigl[ 1/(\tilde{\phi}_1|t^*|)\bigr],
\end{align}
and it follows according to Eqs.\hskip2pt(\ref{EqRhoNum1}) and (\ref{EqRhoNum2}) that
\begin{align}\label{EqTKNum2}
k_{\rm B}T_{\rm K} \propto |t^*|,
\quad 
k_{\rm B}T_{\rm K} =O(|t^*|).
\end{align}
In particular, if Eq.\hskip2pt(\ref{EqRVB-Q2}) is satisfied,
it follows according to Eqs.\hskip2pt(\ref{EqJ_ij}), (\ref{Eq-t*-t}), and (\ref{EqRhoNum2}) that 
\begin{align}\label{EqRhoNum3}
\rho(0)\propto (DU)/(\tilde{\phi}_1 t^2 )\propto D/(\tilde{\phi}_1|J|),
\quad 
\rho(0) =O\bigl[ (DU)/(\tilde{\phi}_1 t^2 )\bigr]
=O\big[D/(\tilde{\phi}_1|J|)\bigr],
\end{align}
and it follows according to Eqs.\hskip2pt(\ref{EqJ_ij}), (\ref{Eq-t*-t}), and (\ref{EqTKNum2}) that 
\begin{align}\label{EqTKNum3}
k_{\rm B}T_{\rm K} \propto t^2/(DU) \propto |J|/D,
\quad
k_{\rm B}T_{\rm K} = O\bigl[ t^2/(DU)\bigr]= O\bigl( |J|/D\bigr).
\end{align}
%
The $k_{\rm B}T_{\rm K}$ is enhanced by the RVB mechanism.
The energy scale $k_{\rm B}T_{\rm K}$ of the Kondo-lattice theory is essentially the same one as the energy scale $|J|/D$ of the RVB theory.

\subsection{On the critical $U_c$ and the self-consistent $1/\tilde{\phi}_1$}
\label{SecCritialUc}
It is a crucial issue whether the self-consistent $\tilde{\phi}_1$ can be defined in a self-consistent solution. This issue is related with another crucial issue whether or not the critical $U_c$, which is defined in a way such that Eq.\hskip2pt(\ref{EqFiniteDelta}) is not satisfied for $U\ge U_c$, exists and, if it exists, whether it is finite or infinite.


In the Anderson model, in general, if Eq.\hskip2pt(\ref{EqFiniteDelta}) is satisfied, then $T_{\rm K}>0\hskip2pt$K; if Eq.\hskip2pt(\ref{EqFiniteDelta}) is not satisfied, then $T_{\rm K}=0\hskip2pt$K.
In general, if $T_{\rm K}>0\hskip2pt$K, the local self-energy can be expanded as in Eq.\hskip2pt(\ref{EqExpansionAM}) and the expansion coefficient $\tilde{\phi}_1$ can be defined; 
if $T_{\rm K}=0\hskip2pt$K, the $\tilde{\phi}_1$ cannot be defined.
The contrapositions of these propositions are as follows:
If the $\tilde{\phi}_1$ can be defined, then $T_{\rm K}>0\hskip2pt$K; if the $\tilde{\phi}_1$ cannot be defined, then $T_{\rm K}=0\hskip2pt$K.
In general, if $T_{\rm K}=0\hskip2pt$K, the low-temperature entropy is so anomalous that the residual entropy is nonzero or the entropy decreases more slowly than the $T$-linear dependence as $T\rightarrow 0\hskip2pt$K.
All of these are also true for the mapped Anderson model.

If the critical $U_c$ exists and it is finite, the self-consistent $\tilde{\phi}_1$ cannot be defined for $U\ge U_c$, so that
$T_{\rm K}=0\hskip2pt$K for $U\ge U_c$. 
If $T_{\rm K}=0\hskip2pt$K, the low-temperature entropy has to be anomalous,
%
%
as discussed above.
Since it is unlikely that a phase with the anomalous low-temperature entropy is stable for a finite $U$ such that $U_c\le U<+\infty$, it is unlikely that the critical $U_c$ is finite. 
This argument implies that if the critical $U_c$ exists for $n=1$ or $n\ne 1$ then the critical $U_c$ has to be infinite. 

If the RVB self-energy is explicitly considered as in this paper, in particular, it is impossible that the critical $U_c$ appears in the range of $|t|\ll U_c<+\infty $, as discussed below.
If the critical $U_c$ is finite and $U_c\gg |t|$, it follows that 
\begin{align}
\bigl[k_{\rm B}T_{\rm K}\bigr]_{U=U_c-0^+} \propto |t^*| =(|t|/\tilde{\phi}_1)
+ 2c_J t^2/(DU_c), \quad
\bigl[k_{\rm B}T_{\rm K} \bigr]_{U=U_c+0^+}=0.
\end{align}
Unless $DU_c$ is infinite, even if $1/\tilde{\phi}_1 \rightarrow 0$ as $U\rightarrow U_c-0^+$, $k_{\rm B}T_{\rm K}$ is discontinuous at the critical $U_c$; thus,
the transition at the $U_c$ is a first-order or discontinuous transition. 
%
The nonzero $k_{\rm B}T_{\rm K}$ means that the energy gain due to the Kondo effect assisted by the RVB mechanism is nonzero, while the zero $k_{\rm B}T_{\rm K}$ means that the energy gain is zero.
Therforer, it is anticipated that the free energy of the phase at $U=U_c-0^+$, whose $k_{\rm B}T_{\rm K}$ is nonzero, is lower than that of the phase at $U=U_c+0^+$, whose $k_{\rm B}T_{\rm K}$ is zero. If this anticipation is true, it contradicts the possibility of the discontinuous transition, so that it contradicts the assumption that the finite $U_c$ exists.  If the critical $U_c$ appears in the range of $U_c \gg |t|$, the critical $U_c$ has to be infinite unless $1/D=0$.

According to Gutzwiller's theory \cite{Gutzwiller1,Gutzwiller2,Gutzwiller3}, which is for the canonical ensemble, 
\begin{subequations}\label{EqGutzwiller}
\begin{align}\label{EqGutzwiller1}
\lim_{U/|t|\rightarrow +\infty} 1/\tilde{\phi}_1
= c_{\rm c}\big|(N/L)-1\bigr|,
\end{align}
where $c_{\rm c}>0$ and $c_{\rm c} =O(1)$, and $N$ is the number of electrons and is an integer. 
Since Gutzwiller's theory is under the single-site approximation, the RVB mechanism is not considered in it.
Either if the RVB mechanism is considered or if it is not considered, it is anticipated that at least
\begin{align}\label{EqGutzwiller2}
1/\tilde{\phi}_1
\ge c_{\rm g}\big|n-1\bigr|,
\end{align}
\end{subequations}
where $c_{\rm g}>0$ and $c_{\rm g}\simeq c_{\rm c}$,  
has to be satisfied for finite or infinite $U/|t|$ in the grand canonical ensemble.
Thus, it is certain that the self-consistent $1/\tilde{\phi}_1$ can be defined at least for $n \ne 1$, even for $n=1\pm 0^+$, and for finite or infinite $U/|t|$.
If $n \ne 1$, the critical $U_c$ does not exist.

According to the analysis in Appendix\hskip2pt\ref{AppAppSumRule2},
if the RVB mechanism is considered and if $n=1$ or $n=1\pm 0^+$, then
\begin{align}\label{EqAsymptotic}
\lim_{U/|t|\rightarrow +\infty}1/\tilde{\phi}_1 
=O\bigl[t^2/\bigl(DU^2\bigr)\bigr].
\end{align}
If $n=1$ or $n=1\pm 0^+$, the critical $U_c$ is infinite.

On there other hand, if $U/|t|\gg1$ and $n=1$, the probability of empty or double occupancy is $O\bigl[t^2/\bigl(DU^2\bigr)\bigr]$.
If $U/|t|\gg1$ and $n$ is very close to unity, it is anticipated that the probability is also $O\bigl[t^2/\bigl(DU^2\bigr)\bigr]$.
If $U/|t|\gg1$ and $n$ is sufficiently different from unity but is still sufficiently close to unity, the probability is proportional to $|n-1|$. Thus, it is anticipated that the probability is as large as $\max\bigl[c_{\rm g}|n-1|, c_{\phi}t^2/\bigl(DU^2\bigr)\bigr]$, where $c_{\phi}>0$ and $c_{\phi}=O(1)$.
It is a reasonable conjecture on the basis of Eqs.\hskip2pt(\ref{EqGutzwiller}) and (\ref{EqAsymptotic}) that 
the spectral weight of the Gutzwiller band, $1/\tilde{\phi}_1$, is proportional to the probability of empty or double occupancy:
\begin{align}\label{EqConjecture2}
1/\tilde{\phi}_1\propto \max\bigl[c_{\rm g}|n-1|, c_{\phi}t^2/\bigl(DU^2\bigr)\bigr].
\end{align}
If this conjecture is true, the electron liquid within the Gutzwiller band is the RVB electron liquid provided that 
\begin{align}
|t|/U \ll 1, \quad 
|n-1| \lesssim  |t|/\bigl(DU\bigr),
\end{align}
are satisfied.
The RVB electron liquid can be stabilized only in the half-filled and almost half-filled cases in the strong-coupling regime.
The conjecture leads to another conjecture: If $|n-1| \ll t^2/\bigl(DU^2\bigr)$, the Gutzwiller band is at the center of the Hubbard gap between the upper and lower Hubbard bands; and 
if $|n-1| \gg t^2/\bigl(DU^2\bigr)$, the Gutzwiller band is at the bottom of the upper Hubbard band or the top of the lower Hubbard band, depending on whether $n>1$ or $n<1$.


\subsection{Metallic conductivity}
\label{SecConductivity}
If $\tilde{\phi}_1\rightarrow+\infty$, the density of states at the chemical potential, $\rho(0)$, is vanishing. It is interesting to examine whether or not the conductivity is vanishing in such a case.

Magnetic impurities are introduced into the Hubbard model:
\begin{align}
\mathcal{H}^\prime &= 
- \sum_{i\sigma\sigma^\prime} J_{i}^{\prime}
\left({\bm \sigma}^{\sigma\sigma^\prime}\cdot{\bm S}_{i}^{\prime}\right)
d_{i\sigma}^\dag d_{i\sigma^\prime}^{\phantom{\dag}},
\end{align}
where ${\bm S}_{i}^{\prime}$ is an impurity spin at the $i$th unit cell.
An ensemble is considered for the exchange-interaction constant $J_i^\prime$.
It is assumed that $J_i^\prime$ is positive, or zero, or negative, and that it is completely random from unit cell to unit cell and from sample to sample:
\begin{align}
\bigl<\hskip-2pt\bigl<J_i^\prime\bigr>\hskip-2pt\bigr> = 0,
\quad
\bigl<\hskip-2pt\bigl<J_i^\prime J_j^\prime\bigr>\hskip-2pt\bigr> = 
\delta_{ij}\overline{\left|J^\prime\right|^2}, 
\quad 
\bigl<\hskip-2pt\bigl<J_i^\prime J_j^\prime J_k^\prime\bigr>\hskip-2pt\bigr> =0 , ~\dots,
\end{align}
where $\bigl<\hskip-2pt\bigl<\cdots\bigr>\hskip-2pt\bigr>$
stands for the ensemble average.
If once the ensemble average is taken, the translational symmetry is restored in the averaged system. Thus, the self-energy due to impurity scatterings is diagonal with respect to the wave number.  
In this subsection, it is assumed that $|J_{i}^\prime|\ll k_{\rm B}T_{\rm K}$.
In the Born approximation, the self-energy  is given by solving self-consistently
\begin{subequations}
\begin{align}
&\hskip20pt
\overline{\Sigma}_{\sigma}(i\varepsilon_l) =
\tilde{\phi}_s^2
S^\prime(S^\prime+1)\overline{\left|J^\prime\right|^2}
\frac1{L}\sum_{\bm k}\frac1{\tilde{\phi}_1}\overline{g}_{\sigma}(i\varepsilon_l,{\bm k}),
\\
&\overline{g}_{\sigma}(i\varepsilon_l,{\bm k})=\frac1{i\varepsilon_l +\mu^* -\xi({\bm k}) - \tilde{\gamma}_{\rm K}(i\varepsilon_l) - (1/\tilde{\phi}_1)\overline{\Sigma}_\sigma(i\varepsilon_l) },
\end{align}
\end{subequations}
where $S^\prime$ is the magnitude of the impurity spins, 
$\overline{\Sigma}_{\sigma}(i\varepsilon_l)$ is the ensemble-averaged self-energy, which corresponds to 
$\Gamma(i\varepsilon_l)$ introduced in \S\ref{SecModel}, 
and $\overline{g}_{\sigma}(i\varepsilon_l,{\bm k})$ is the ensemble-averaged one of $g_{\sigma}(i\varepsilon_l,{\bm k})$, which is defined by Eq.\hskip2pt(\ref{EqGr-KL}).
Since $\tilde{W}_s=\tilde{\phi}_s/\tilde{\phi}_1=O(1)$, as shown in Eq.\hskip2pt(\ref{Eq3W}), $(1/\tilde{\phi}_1)\overline{\Sigma}_\sigma(i\varepsilon_l)$ is of the zeroth order in $1/\tilde{\phi}_1$.
According to Eq.\hskip2pt(\ref{EqD-Pi}), $\tilde{\phi}_1/[\pi\tilde{\Delta}(0)]$ is of the zeroth order in $1/\tilde{\phi}_1$,
so that $\tilde{\gamma}_{\rm K}(i\varepsilon_l)$ is also of the zeroth order in $1/\tilde{\phi}_1$.

According to the Kubo formula \cite{kubo},
the electrical conductivity is given by
\begin{align}\label{EqKubo}
\sigma_{xx}(\omega) = 
\frac{\hbar}{{i}\omega}\bigl[
K_{xx}(\omega+{i}0)-K_{xx}(0)\bigr],
\end{align}
where $K_{xx}(\omega+{i}0)$ is the retarded one of 
%
\begin{align}\label{EqKXXT}
K_{xx}({i}\omega_l) &= 
\frac1{La^D}\int_{0}^{1/(k_{\rm B}T)} \hskip-10pt d\tau
e^{{i}\omega_l\tau}\left< e^{\tau\mathcal{H}}
\hskip1pt \hat{j}_x \hskip1pt e^{-\tau\mathcal{H}} \hskip1pt \hat{j}_x\right> =
\frac{e^2}{\hbar^2}\frac{(2t)^2}{D a^{D-2}}
 \Pi_{xx}({i}\omega_l),
\end{align}
where 
$\hat{j}_x$ is the first or $x$ component of the current operator defined by
\begin{align}
\hat{\bm j} &=
\sum_{{\bm k}\sigma} {\bm j}({\bm k})\hat{n}_{\bm k} ,
\quad
{\bm j}({\bm k}) = 
-\frac{e}{\hbar}
\frac{\partial\hskip8pt}{\partial{\bm k}}E({\bm k}),
\quad
\hat{n}_{{\bm k}\sigma} = 
\frac1{L}
\sum_{ij} e^{i{\bm k}\cdot\left({\bm R}_i-{\bm R}_j\right)} d_{i\sigma}^\dag d_{j\sigma},
\end{align}
and $\Pi_{xx}({i}\omega_l)$ is defined by
\begin{align}
\Pi_{xx}({i}\omega_l) &=
\frac1{L}\sum_{{\bm k}{\bm p}}\sum_{\sigma\sigma^\prime}
\sin(k_x a)\sin(p_x a)
\int_{0}^{1/(k_{\rm B}T)} \hskip-10pt d\tau
e^{{i}\omega_l\tau} 
%
\left<e^{\tau\mathcal{H}}\hskip1pt \hat{n}_{{\bm k}\sigma}\hskip1pt e^{-\tau\mathcal{H}}
\hskip1pt \hat{n}_{{\bm p}\sigma^\prime}\right>,
\end{align}
%
where $k_x=k_1$ and $p_x=p_1$.
The current vertex $j_x({\bm k})$ has to be consistently renormalized with $\overline{\Sigma}_{\sigma}(i\varepsilon_l)$ and $\Delta\Sigma_\sigma^{\rm (RVB)}(i\varepsilon_l,{\bm k})$
in order to satisfy the Ward relation \cite{ward}. The vertex correction due to impurity scatterings vanishes in the Born approximation, while
the ladder vertex of the first order in the $J_s(i\omega_l,{\bm q})$ has to be considered.
If Eqs.\hskip2pt(\ref{EqJs2}) and (\ref{EqJsKP}) are used, it follows that 
\begin{align}\label{EqUpperPiJ}
& \hskip30pt 
\Pi_{xx}({i}\omega_l) =
\frac{1}{\tilde{\phi}_1^2}
\frac{2\pi_{xx}({i}\omega_l)}
{\displaystyle 1 + (3J/2D)\tilde{W}_s^2\pi_{xx}({i}\omega_l)}, 
\\ \label{EqLowerPiJD} &
\pi_{xx}({i}\omega_l) = 
- \frac{k_{\rm B}T}{L}\sum_{n{\bm k}}
\sin^2(k_x a)\overline{g}_\sigma(i\varepsilon_l,{\bm k})\overline{g}_\sigma(i\varepsilon_l+{i}\omega_l,{\bm k}).
\end{align}
According to Eq.\hskip2pt(\ref{EqKubo}), the $\omega_l$-linear term of $K_{xx}(i\omega_l)$ contributes to the static conductivity $\sigma_{xx}(0)$, so that the $\omega_l$-linear term of $\Pi_{xx}(i\omega_l)$ contributes to it. 
It follows that
\begin{align}\label{EqTwoVerJ1}
\frac{d \Pi_{xx}({i}\omega_l)}
{d({i}\omega_l)} &=
\frac{2}{\tilde{\phi}_1^2} \frac1
{\displaystyle \left[1 + (3J/2D)\tilde{W}_s^2\pi_{xx}({i}\omega_l)\right]^2}
\frac{d \pi_{xx}({i}\omega_l) }{d({i}\omega_l)}.
\end{align}
According to Eq.~(\ref{EqXiPi}), it follows that
\begin{align}\label{EqWardtdtd*}
\tilde{\phi}_1\left[1 + (3J/2D) \tilde{W}_s^2\pi_{xx}(0)\right] &= 
t/t^*.
\end{align}
If Eqs.\hskip2pt(\ref{EqTwoVerJ1}) and (\ref{EqWardtdtd*}) are used, the static conductivity is simply given by
\begin{align}\label{EqTwoVer2}
\sigma_{xx}(0) &=
\frac{e^2}{\hbar^2}\frac{(2t^*)^2}{Da^{D-2}}
S_{xx}(0),
\\ \label{EqPJ1} 
S_{xx}(0) &=
\lim_{\omega\rightarrow 0}
\frac{2\hbar}{{i}\omega} \left[
\pi_{xx}(\omega+{i}0) - \pi_{xx}(0)
\right]
\nonumber \\ &=
\frac{2\hbar}{\pi L} \sum_{\bm k}\sin^2(k_xa)
\int_{-\infty}^{+\infty} \hskip-10pt d\varepsilon 
\left[-\frac{f(\varepsilon)}{d \varepsilon }\right]
\Bigl[{\rm Im} \hskip1pt    
\overline{g}_\sigma(\varepsilon+{\rm i}0,{\bm k})\Bigr]^2.
\end{align}
If $2\sin^2(k_xa) = 1-\cos(2 k_xa)$ is used and the term that includes $\cos(2k_xa)$ is ignored, then
\begin{align}\label{EqPJ2}
S_{xx}(0)  & =
\frac{\hbar}{\pi} \hskip-2pt \int \hskip-2pt dE \hskip-2pt \left[\tilde{\phi}_1\rho(E)\right]\hskip-2pt
\int_{-\infty}^{+\infty} \hskip-10pt d\varepsilon \hskip-2pt
\left[-\frac{f(\varepsilon)}{d \varepsilon }\right]
\hskip-3pt \left[{\rm Im} \hskip1pt    
\frac1{\varepsilon-E-\tilde{\gamma}_{\rm K}(\varepsilon+i0) -\bigl(1/\tilde{\phi}_1\bigr)\overline{\Sigma}_\sigma(\varepsilon+i0) }\right]^2.
\end{align}
%
If $T\ll T_{\rm K}$ or $T_{\rm N}<T\ll T_{\rm K}$, depending on whether $T_{\rm N}$ does not exists or it exists,
and if 
\begin{align}
& i\frac{\hbar}{2\tau_{\rm K}} =
\int_{-\infty}^{+\infty}\hskip-10pt d\varepsilon
\left[-\frac{df(\varepsilon)}{d\varepsilon} \right]
\bigl[-\tilde{\gamma}_{\rm K}(\varepsilon+i0)\bigr]
=i\left(\frac{\pi^2}{3}\tilde{W}_{21}+\tilde{W}_{22}\right)
\frac{\tilde{\phi}_1}{\pi\tilde{\Delta}(0)}\left(k_{\rm B}T\right)^2,
\\ & \hskip70pt
\frac{\hbar}{2\tau_s} =
\int_{-\infty}^{+\infty}\hskip-10pt d\varepsilon
\left[-\frac{df(\varepsilon)}{d\varepsilon} \right]
\bigl[-(1/\tilde{\phi}_1){\rm Im}\overline{\Sigma}_\sigma(\varepsilon+i0)\bigr],
\end{align}
are approximately used for $-\tilde{\gamma}_{\rm K}(\varepsilon+i0)$ and $-(1/\tilde{\phi}_1){\rm Im}\overline{\Sigma}_\sigma(\varepsilon+i0)$, respectively,
and if the energy dependences of $\rho(E)$ and $(1/\tilde{\phi}_1){\rm Re} \overline{\Sigma}_\sigma(\varepsilon+i0)$ are ignored, 
then it follows that
\begin{align}
\sigma_{xx}(0) &=
\frac{e^2}{\hbar^2}\frac{8|t^*|^2}{Da^{D-2}}
\frac{\tilde{\phi}_1\rho(0)}{(1/\tau_{\rm K})+ (1/\tau_s)}.
\end{align}

%
%
%
%

Since $|t^*|=O(|J|/D)$ and $\tilde{\phi}_1\rho(0)=O(1/|t^*|)$,
the conductivity is of the zeroth order in $1/\tilde{\phi}_1$. 
Thus, it does not vanish 
even in the limit of $\tilde{\phi}_1\rightarrow +\infty$, although $\rho(0)|t| \rightarrow 0$ as $\tilde{\phi}_1\rightarrow +\infty$.
In a clean system, $\hbar/\tau_s=0$. 
If no symmetry is broken even at $T=0\hskip2pt$K in the clean system, then $1/\tau_{\rm K}\rightarrow 0$ as $T\rightarrow 0\hskip2pt$K, so that the conductivity diverges as $T\rightarrow 0\hskip2pt$K.
The RVB electron liquid shows a metallic conductivity at least under the Born approximation,%
\footnote{If impurity scatterings are rigorously treated, even if they are weak, the conductivity of the RVB electron liquid in one and two dimensions has to be vanishing as $T\rightarrow0\hskip2pt$K because of the Anderson localization \cite{abrahams}.}
although it is almost a spin liquid in the sense that the density of states at the chemical potential, $\rho(0)$, is almost vanishing.

%

\section{Adiabatic Continuation}
\label{SecAdiabatic}
In the Heisenberg limit of $U/|t|\rightarrow +\infty$, with $J = -4t^2/U$ being kept constant, 
the half-filled Hubbard model is reduced to
\begin{align}\label{EqHeisenberg}
\mathcal{H}_S &=
\frac1{2}L(\epsilon_d-\mu) + \mathcal{P}^{-1}\biggl[
- \frac1{2}\frac{J}{D} \sum_{ij}\delta_{\left<ij\right>}
\left({\bm S}_i\cdot{\bm  S}_j\right)
- 2 \sum_{i} J_{i}^{\prime}\left({\bm S}_i\cdot{\bm S}_i^\prime\right)
\biggr]\mathcal{P},
\end{align}
where $\epsilon_d-\mu=-U/2$, the impurity term considered in \S\ref{SecConductivity} is included, and
\begin{align}
\mathcal{P} &= \prod_{i}\left(1-n_{i\uparrow}n_{i\downarrow}\right)
\left(n_{i\uparrow} + n_{i\downarrow}\right),
\quad
{\bm S}_i = 
\frac1{2}\sum_{\sigma\sigma^\prime}{\bm \sigma}^{\sigma\sigma^\prime}
d_{i\sigma}^\dag d_{i\sigma^\prime}^{\phantom{\dag}}.
\end{align}
Because of  the projection operator $\mathcal{P}$, empty and double occupancies are completely excluded.
Since ${\bm S}_i$'s satisfy the commutation relation for spin within the constrained Hilbert subspace where no empty nor double occupancy is allowed, Eq.\hskip2pt(\ref{EqHeisenberg}) is none other than the Heisenberg model.

It is easy to confirm that
\begin{align} 
\bigl[\mathcal{H}_S, n_{i\uparrow}+n_{i\downarrow} \bigr]= 0, \quad \bigl[\mathcal{H}, n_{i\uparrow}+n_{i\downarrow} \bigr]\ne 0,
\end{align}
%
for any $i$th site.
The local gauge symmetry exists in the Heisenberg model, but it does not in the Hubbard model; and
the conductivity of the Heisenberg model is absolutely zero, but the conductivity of the Hubbard model with no impurity can be divergent at $T=0\hskip2pt$K even in the Heisenberg limit, unless no symmetry is broken even at $T=0\hskip2pt$K, as studied in \S\ref{SecConductivity}.
This situation for the Heisenberg and Hubbard models is similar to  the situation for the $s$-$d$ and Anderson model.
In the $s$-$d$ limit, the Anderson model is reduced to the $s$-$d$ model;
the local gauge symmetry exists in the $s$-$d$ model, but it does not exist in the  Anderson model.

The local gauge symmetry is a peculiar symmetry.
The local gauge symmetry  cannot be spontaneously broken \cite{elitzur}.
In the reduction of the Hubbard and Anderson models into the Heisenberg and $s$-$d$ models, the local gauge symmetry  is {\it not spontaneously} restored but is {\it forced} to be restored by the constraint of the Hilbert space within the subspace where no empty nor double occupancy is allowed.
One of the most peculiar features of the local gauge symmetry is that
the difference of it
is irrelevant to the adiabatic continuation between the local electron liquid in the Anderson model and the local spin liquid in the $s$-$d$ model.

The strength of magnetic impurities can be used as an adiabatic parameter. 
Here, it is assumed that $-\infty< J_{i}^{\prime} <+\infty$ and $0<\overline{\left|J^\prime\right|^2}<+\infty$.
Clean and dirty limits are defined by the limit of $\overline{\left|J^\prime\right|^2}\rightarrow 0$ and the limit of $\overline{\left|J^\prime\right|^2}\rightarrow +\infty$, respectively.
In the dirty-limit Hubbard model, an electron is localized almost within a unit cell, so that the local gauge symmetry is almost {\it restored} and the conductivity is almost zero.
Therefore, it is certain that every physical property of the dirty-limit  Hubbard model in the Heisenberg limit is the same as that of the dirty-limit Heisenberg model. Thus, the electron state in the dirty-limit Hubbard model  in the Heisenberg limit and the spin state in the dirty-limit Heisenberg model are adiabatically connected to each other.

According to the scaling theory for the Anderson localization  \cite{abrahams},
there is no critical point between metallic and insulating phases, or between itinerant and localized states, or between the clean and dirty limits; and there is no lower limit of the metallic conductivity nor no minimum metallic conductivity.
Therefore, the RVB electron liquid in the clean-limit Hubbard model is adiabatically connected to the electron state in the dirty-limit Hubbard model.
It is obvious that the RVB spin liquid in the clean-limit Heisenberg model is adiabatically connected to the spin state in the dirty-limit Heisenberg model. Thus, the RVB electron liquid in the clean-limit Hubbard model and the RVB spin liquid in the clean-limit Heisenberg model are adiabatically connected to each other.
The difference of the local gauge symmetry is also irrelevant to the adiabatic continuation between the RVB electron liquid in the Hubbard model and the RVB spin liquid in the Heisenberg model.

\section{Discussion}
\label{SecDiscussion}
The Hubbard model in one dimension is particular, because 
no symmetry can be broken \cite{mermin}, and because the Bethe-ansatz exact solution was given \cite{lieb-wu}.
Since the Bethe-ansatz solution is for the canonical ensemble, 
the electron number $N$ is an integer.
According to the Bethe-ansatz solution,
the ground state is not the Mott insulator at least for any non-half-filling, i.e., for any $N\ne L$ and even for $N=L\pm 1$.
In the grand canonical ensemble, on the other hand, the averaged electron number $\left<\mathcal{N}\right>$ can be an irrational. 
If a critical $\delta N_c$ is defined such that the ground state is a metal for $|\left<\mathcal{N}\right>-L|> \delta N_c$ while it is the Mott insulator for $|\left<\mathcal{N}\right>-L|\le \delta N_c$, 
it is anticipated that the critical $\delta N_c$, if it exists, is within the range of $0\le \delta N_c<1$.
In the thermodynamic limit, the electron density is defined by $n=\lim_{L\rightarrow +\infty}\left<\mathcal{N}\right>/L$. 
At least if $\delta N_c/L \rightarrow 0$ as $L\rightarrow +\infty$, the ground state is a metal 
for any $n\ne 1$, even for $n=1\pm 0^+$, in the thermodynamic limit. 
The Hubbard model with $n=1\pm 0^+$ is none other than the {\it half-filled} Hubbard model.
The ground state of the {\it half-filled} Hubbard model is the Tomonaga-Luttinger electron liquid \cite{solyom,Luttinger-liq} and is also the RVB electron liquid.
If $\delta N_c/L \rightarrow 0$ as $L\rightarrow +\infty$, effects of $\delta N_c/L$ being different from zero can be ignored in the thermodynamic limit, so that the RVB electron liquid in the {\it half-filled} Hubbard model with $n=1\pm 0^+$ is adiabatically connected to the RVB spin liquid in the Heisenberg model.

It is assumed that $\delta N_c/L \rightarrow 0$ as $L\rightarrow +\infty$, and that the chemical potential $\mu$ is one such that $n=1\pm 0^+$ for it,
and that $U/|t|\gg 1$ or $U/|t|\rightarrow +\infty$
with $J=-4t^2/U$ being kept constant.
If the conjecture of Eq.\hskip2pt(\ref{EqConjecture2}) is true or not, $1/\tilde{\phi}_1$ is almost infinitesimal but still nonzero. Since $1/\tilde{\phi}_1$ can be defined, $k_{\rm B}T_{\rm K}$ is nonzero: $k_{\rm B}T_{\rm K}\propto |J|$ and $k_{\rm B}T_{\rm K}=O(|J|)$. Therefore, the ground state is a metal, provided that no impurity exists and the Anderson localization does not occur, as studied in \S\ref{SecConductivity}. 
According to the Fermi-surface sum rule \cite{Luttinger1,Luttinger2}, the Fermi surface or point is given by
\begin{align}\label{EqFS-1D}
k_{\rm F} = \pm \bigl[\pi/(2a)\bigr](1+\delta n),
\end{align}
where $\delta n = n- 1 = \pm0^+$. If no multisite term but the RVB self-energy is considered, the single-particle excitation spectrum is given by
\begin{align}
\xi(k) &=
c_J J \cos(ka),
\end{align}
where $c_J\simeq 1$.
The $\mu^*$, which is defined by Eq.\hskip2pt(\ref{EqGreenKL3}), can be determined by the Fermi-surface sum rule: $\mu^*=c_J|J|\sin (\pi \delta n/2)$. 
Since $\delta n = \pm0^+$, $\mu^*=\pm 0^+$.
Then, a pair excitation spectrum $\omega(q)$ is given by
\begin{align}
\omega(q) = c_J |J|\left\{ \cos[(k+q)a] - \cos(ka)\right\},
\end{align}
where $k$ and $q$ are restricted to $\cos[(k+q)a]-\mu^*>0$ and $\cos(ka)-\mu^*<0$.
Then, 
\begin{align}
c_J |J\sin(qa)| \le \omega(q) \le 2c_J |J\sin(qa/2)|.
\end{align}
On the other hand, the spectral weight of the Gutzwiller band is almost vanishing because $1/\tilde{\phi}_1$ is almost infinitesimal. 
Furthermore, 
if $U/|t|\gg 1$ or $U/|t|\rightarrow +\infty$ and if $\delta n=\pm 0^+\hskip-1pt$, 
low-energy quantum charge fluctuations are much suppressed and vanishing. Thus, the spectral weight in the charge channel of the pair excitation $\omega(q)$ is vanishingly small, i.e., the pair excitation $\omega(q)$ is almost a spin excitation.
This spin-excitation spectrum in the Hubbard model in one dimension resembles that in the Heisenberg model in one dimension \cite{cloiseaux}.


On the basis of the criterion that if a gap opens in the spectrum of adding or removing a single electron in the canonical ensemble then the ground state has to be an insulator, as well as the exact result of the Bethe-ansatz solution that the gap opens for any nonzero $U$,
Lieb and Wu proposed that the half-filled ground state is the Mott insulator for any nonzero $U$ \cite{lieb-wu}. 
On the other hand, according to the Bethe-ansatz solution, the residual entropy per unit cell is zero or vanishing as $L\rightarrow +\infty$. According to the previous paper \cite{FJO-MottIns}, if a complete gap opens in the self-consistent $\rho(\varepsilon)$, the residual entropy is $k_{\rm B}\ln 2$ per unit cell;
even if the RVB mechanism is considered, this conclusion does not change.
On the basis of these two results on the residual entropy, it can also be proposed that the half-filled ground state cannot be an insulator with a complete gap open.
The criterion used by Lieb and Wu in order to conclude that the ground state is the Mott insulator is not a sufficient condition for the ground state being an insulator but a necessary condition for it; e.g., a similar gap opens in a metallic fine particle because of the long-range Coulomb interaction.
Thus, it is desirable to critically reexamine whether the half-filled ground state of the Hubbard model in one dimension is the Mott insulator or a metal.
If it is the Mott insulator, the critical $U_c$, which is defined on the basis of Eq.\hskip2pt(\ref{EqFiniteDelta}), is zero for $D=1$ and $\left<\mathcal{N}\right>=L$, or $n=1$; and
it has to be determined how large or small the critical $\delta N_c$, which is defined above, and $\lim_{L\rightarrow+\infty}(\delta N_c/L)$ are.
If it is a metal, the argument above for $n=1\pm 0^+$ is valid for the exactly half-filled case of $n=1$.

On the basis of the adiabatic continuation, it is anticipated that physical properties resemble each other between the Hubbard and Heisenberg models even in one dimension, except for physical properties that are related to the itineracy of electrons.
It has already been proposed that the spin liquid in one dimension can be described as the Tomonaga-Luttinger spin liquid \cite{TL1,TL2,TL3}.
{\it The existence of charge fluctuations is not a necessary condition for the Fermi or Tomonaga-Luttinger liquid, i.e., a normal or anomalous Fermi liquid to be stabilized}, as discussed in \S\ref{SecIntroduction}.
According to the study of the present paper,
this proposal is none other than the proposal that the Tomonaga-Luttinger electron and spin liquids are also the RVB electron and spin liquids, respectively, and that they are adiabatically connected to each other.
The proposal that the ground state of the Heisenberg model is the Tomonaga-Luttinger spin liquid never contradicts the proposal that the exactly half-filled ground state of the Hubbard model is the Tomonaga-Luttinger electron liquid and not the Mott insulator.

The half-filled Hubbard model on the square lattice is also particular, because no symmetry can be broken at a nonzero $T$ \cite{mermin}, and because $\rho_0(\varepsilon)$ diverges logarithmically as $\varepsilon\rightarrow 0$.
It follows that
\begin{align}
1/\chi_s(0,{\bm q};T) &
= [1/\tilde{\chi}_s(0;T)]-(1/4)I_s(0,{\bm q};T) 
\nonumber \\ & 
= k_{\rm B}T_{\rm K}(T) 
\Bigl\{1-O\Bigl[[T/T_{\rm K}(T)]^2\Bigr]\Bigr\}
- (1/4)I_s(0,{\bm q}; T),
\end{align}
where the Fermi-liquid relation studied in \S\ref{SecNormalFL} is used. 
In the temperature range defined by Eq.\hskip2pt(\ref{EqTregion}), but with $\delta T$ being sufficiently large, the logarithmic divergence of $\tilde{\phi}_1\tilde{\rho}(\varepsilon)$ as $\varepsilon\rightarrow 0$ is suppressed by the imaginary part of the self-energy.
Then, $T_{\rm K}(T) \simeq |J|/(2k_{\rm B})$ or $T_{\rm K}(T) =O\bigl[|J|/(2k_{\rm B})\bigr]$.
On the other hand, if $T$ is very low such that $T\ll |J|/(2k_{\rm B})$, the increase or divergence of $\tilde{\phi}_1\tilde{\rho}(\varepsilon)$ as $\varepsilon\rightarrow 0$ is substantial,
so that $T_{\rm K}(T) \rightarrow 0\hskip2pt$K and $1/\tilde{\chi}_s(0;T)\rightarrow 0$ as $T\rightarrow 0\hskip2pt$K.%
\footnote{Since the vanishing $T_{\rm K}(T)$ as $T\rightarrow 0\hskip2pt$K is simply because of the divergence of $\tilde{\phi}_1\rho(\varepsilon)$, it never means that the condensation energy of the RVB electron liquid is also vanishing as $T\rightarrow 0\hskip2pt$K. The condensation energy is $O(|J|/2)$ per unit cell.}
%

Since $1/\tilde{\chi}_s(0;T)\rightarrow 0$ as $T\rightarrow 0\hskip2pt$K and the superexchange interaction $J_s(0,{\bm q})$ is maximum at ${\bm q}={\bm Q}$, where ${\bm Q}=(\pm 1, \pm 1)(\pi/a)$, the ground state un the square lattice is the N\'{e}el state whose ordering wave number is ${\bm Q}$ or the RVB electron liquid very close to the N\'{e}el state.
If $T$ is sufficiently lower than $|J|/(2k_{\rm B})$, it is plausible that such a low-$T$ phase is in the critical region.
%
If the low-$T$ phase is really in the critical region of antiferromagnetism, it is anticipated that an anomaly appears in the uniform susceptibility, as discussed below.
It follows that
\begin{subequations}\label{Eq1/Chi-2}
\begin{align}
& \hskip3pt 
1/\chi_s(0,0; T) =
1/\chi_s(0,{\bm Q};T) +
(1/4)\bigl[\Delta_s + \Delta_Q(T) + \Delta_\Gamma(T)\bigr],
\\ &
\Delta_s = J_s(0,{\bm Q})- J_s(0,0), \quad
\Delta_Q(T)  = J_Q(0,{\bm Q};T)- J_Q(0,0;T), \quad
\\ & \hskip65pt
\Delta_\Gamma(T)  = \Lambda_\Gamma(0,{\bm Q};T)- \Lambda_\Gamma(0,0;T). 
\end{align}
\end{subequations}
%
The $J_s(0,{\bm q})$ is antiferromagnetic, i.e., $J_s(0,{\bm Q})>0$ and $J_s(0,0)<0$; and $\Delta_s = 4|J|$, which is almost independent of $T$.
According to previous papers \cite{FJO-supJ,miyai,FJO-review}, 
if the density of states has a sharp peak at one of the band edges and the chemical potential is in the vicinity of the peak position, $J_Q(0,{\bm q};T)$ is strongly ferromagnetic, i.e., $J_Q(0,0;T)$ is positively large and $J_Q(0,0;T)$ increases as $T\rightarrow 0\hskip2pt$K; if the Fermi surface shows a sharp nesting, it is strongly antiferromagnetic, i.e., $J_Q(0,{\bm Q}_{\rm N};T)$, where ${\bm Q}_{\rm N}$ is a nesting wave number, is positively large and $J_Q(0,{\bm Q}_{\rm N};T)$ increases as $T\rightarrow 0\hskip2pt$K.%
\footnote{Since $\Lambda(0,{\bm q};T)$ is of higher order in $1/D$, its $T$ dependence cannot be responsible for the Curie-Weiss $T$ dependence in sufficiently high dimensions. Thus, only the $T$ dependence of $J_Q(0,0;T)$ or $J_Q(0,{\bm Q}_{\rm N};T)$ can be responsible for the Curie-Weiss $T$ dependence of itinerant-electron magnetism in sufficiently high dimensions; on the other hand, only the $T$ dependence of $\tilde{\chi}_s(0;T)$ can be responsible for the Curie-Weiss $T$ dependence of local-moment magnetism in sufficiently high dimensions, as discussed in \S\ref{SecDiscussion}.}
In the half-filled Hubbard model on the square lattice, the Fermi surface shows a sharp nesting for ${\bm Q}=(\pm 1, \pm 1)(\pi/a)$, so that $J_Q(0,{\bm Q};T)$ is positively large at a sufficiently low $T$ and $J_Q(0,{\bm Q};T)$ increases as $T\rightarrow 0\hskip2pt$K; and the density of states has a logarithmic peak at the chemical potential, so that $J_Q(0,0;T)$ is also positive at a sufficiently low $T$ and $J_Q(0,0;T)$ increases as $T\rightarrow 0\hskip2pt$K.
Since the peak of the density of states is at the band center and 
since the nesting effect is larger or stronger than the logarithmic-peak effect, 
$\Delta_Q(T) >0$
and the $T$ dependence of $J_Q(0,{\bm Q};T)$ is much stronger than that of $J_Q(0,0;T)$. 
Thus, the $T$ dependence of $\Delta_Q(T)$ is large;  $\Delta_Q(T)$ increases as $T\rightarrow 0\hskip2pt$K.
The N\'{e}el temperature $T_{\rm N}$ cannot be nonzero because of critical fluctuations or $\Lambda(0,{\bm q};T)$, which means that the $T$ dependence of $\Lambda(0,{\bm q};T)$ is large.
In general, the ${\bm q}$ dependence of the mode-mode coupling term is small. Then, the ${\bm q}$ dependence of $\Lambda(0,{\bm q};T)$ is small, and $\Delta_\Gamma(T)$ is therefore small, and the absolute, not relative, $T$ dependence of the small $\Delta_\Gamma(T)$ is also small.
In the critical region, in general, the $T$ dependence of $1/\chi_s(0,{\bm Q};T)$ deviates from the Curie-Weiss $T$ dependence.
If the low-$T$ phase is really in the critical region, $1/\chi_s(0,{\bm Q};T)\simeq 0$ has to be satisfied in it.
If the $T$ dependence $\Delta_\Gamma(T)$ is really small, therefore, it is anticipated on the basis of Eq.\hskip2pt(\ref{Eq1/Chi-2}) that the $T$ dependence of $1/\chi_s(0,0; T)$ resembles the $T$ dependence of $\Delta_Q(T)$ in the critical region.
Since $\Delta_Q(T)$ increases as $T\rightarrow 0\hskip2pt$K, it is anticipated that $1/\chi_s(0,0; T)$ increases as $T\rightarrow 0\hskip2pt$K.
It is interesting to examine whether or not $1/\chi_s(0,0; T)$ increases or $\chi_s(0,0; T)$ decreases as $T\rightarrow 0\hskip2pt$K in the half-filled Hubbard model on the square lattice.
If the decrease of $\chi_s(0,0; T)$ as $T\rightarrow 0\hskip2pt$K is true, it
corresponds to the suppression of the static and uniform susceptibility as $T\rightarrow 0\hskip2pt$K in the Heisenberg model on the square lattice \cite{HZ2-1,HZ2-2}.

It is straightforward to extend the study in this paper to the triangular lattice.
If $T>0\hskip2pt$K, no symmetry can be broken \cite{mermin}.
The electron state at $0\hskip2pt{\rm K}<T \ll T_{\rm K}$ or $0\hskip2pt{\rm K}< T \ll |J|/(2k_{\rm B})$ in the Hubbard model is a frustrated electron liquid in the sense that no symmetry is broken in it, and the spin state at $0\hskip2pt{\rm K}< T \ll |J|/(2k_{\rm B})$ in the Heisenberg model is the RVB spin liquid proposed by Fazekas and Anderson \cite{fazekas}.
We propose that the frustrated electron liquid in the Hubbard model is none other than the RVB electron liquid, and that the RVB electron and spin liquids in the Hubbard and Heisenberg model on the triangular lattice are adiabatically connected to each other.

In three dimensions and higher, it is possible that $T_{\rm N} \ll |J|/(Dk_{\rm B})$ because of frustration and quasi-one or quasi-two dimensionality in the Hubbard and Heisenberg models, and
the deviation from the half-filling in the Hubbard model, and so on.
It is interesting to study how magnetic properties at $T_{\rm N}< T \ll |J|/(Dk_{\rm B})$ resemble each other between an electron liquid in the Hubbard model in the strong-coupling regime, which is the RVB electron liquid, and a spin liquid in the Heisenberg model, which is the RVB spin liquid.


If $D\rightarrow +\infty$ and $n=1$, $k_{\rm B}T_{\rm K}\propto |J|/D\rightarrow 0$ and $\Lambda(0,{\bm q})\rightarrow 0$. If $T/T_{\rm K}\rightarrow +\infty$,
$J_Q(0,{\bm q})\rightarrow 0$ and $\Lambda(0,{\bm q})\rightarrow 0$.
If $T\gg T_{\rm K}$ and $n= 1$,
the static susceptibility of the Anderson model is given by
\begin{align}
\tilde{\chi}_s(0) = 1/(k_{\rm B}T)- O\bigl[T_{\rm K}/(k_{\rm B}T^2)\bigr].
\end{align}
Thus, if $D\rightarrow +\infty$ and $n=1$, the static susceptibility of the Hubbard model is given by
\begin{subequations}\label{EqCW-Law}
\begin{align}\label{EqCW-Law1}
\chi_s(0,{\bm q}) = \frac{\tilde{\chi}_s(0)}{1-(1/4)J_s(0,{\bm q})\tilde{\chi}_s(0)}
= \frac1{k_{\rm B}T- (1/4)J_s(0,{\bm q}) }.
\end{align}
This $\chi_s(0,{\bm q})$ agrees with the susceptibility in the mean-field approximation for the Heisenberg model.
The N\'{e}el temperature is given by
\begin{align}\label{EqCW-Law2}
T_{\rm N} = J_s(0,{\bm Q})/(4k_{\rm B})
=|J|/(2k_{\rm B}),
\end{align}
\end{subequations}
where ${\bm Q}=(\pm 1, \pm 1, \cdots, \pm 1)(\pi/a)$.
This $T_{\rm N}$ also agrees with the N\'{e}el temperature in the mean-field approximation for the Heisenberg model. 
Since the mean-field approximation becomes rigorous in the limit of $D\rightarrow +\infty$, these agreements are quite reasonable.

If $T_{\rm N}\ll T_{\rm K}$ then itinerant-electron magnetism appears at $T\le T_{\rm N}$, while
if $T_{\rm N}\gg T_{\rm K}$ then local-moment magnetism appears at $T\le  T_{\rm N}$. 
Thus, itinerant-electron magnetism and local-moment magnetism are characterized by $T_{\rm N}\ll T_{\rm K}$ and $T_{\rm N}\gg T_{\rm K}$, respectively, and they are adiabatically connected to each other.
If $T_{\rm N}\gg T_{\rm K}$, a paramagnetic phase at $T\gtrsim T_{\rm N}$ is none other than the paramagnetic phase of the insulator proposed by Slater \cite{slater}, or the antiferromagnetic type of the Mott insulator. 
The entropy of the paramagnetic phase is as large as $k_{\rm B}\ln 2$ per unit cell, and the static susceptibility of it obeys Eq.\hskip2pt(\ref{EqCW-Law}), at least approximately.
In particular, if $D\rightarrow +\infty$ and $n=1$, then $T_{\rm K}/T_{\rm N}\rightarrow 0$, as discussed above. Thus, magnetism for $D\rightarrow +\infty$ and $n=1$ is a typical one of local-moment magnetism; and the electron state for $D\rightarrow +\infty$ and $n=1$ is a typical one of the antiferromagnetic type of the Mott insulator.
It is obvious that the local-moment states in the Hubbard and Heisenberg models are adiabatically connected to each other.

The RVB electron liquid studied in this paper is none other than a {\it normal} state in order to study possible low-temperature ordered phases such as the N\'{e}el state of itinerant-electron magnetism, which is of the zeroth order in $1/D$, an anisotropic superconducting state, which is of higher order in $1/D$, and so on in the strong coupling regime defined by $U/|t|\gg 1$ and in the half-filled or almost half-filled case. 
It is plausible that the {\it normal} state proposed by Anderson \cite{Anderson-SC} is none other than the RVB electron liquid studied in this paper.
The study of this paper confirms the relevance of theory of high-temperature superconductivity in cuprate oxides based on the Kondo-lattice theory \cite{FJO-SC1,FJO-SC2}.
If the half-filled ground state in one dimension is the Mott insulator, as was proposed by Lieb and Wu \cite{lieb-wu}, it is anticipated that the RVB electron liquid studied in this paper can also be used as a {\it normal} state in order to study the Mott insulator in one dimension as a possible low-temperature phase. It is interesting to elucidate what effect is responsible for a complete gap to open in the Mott insulator in one dimension; within a preliminary study, however, there is no evidence that the self-energy can be so anomalous that it can make a complete gap open.

\section{Conclusion}
\label{SecConclusion}
The resonating-valence-bond (RVB) electron liquid 
in the half-filled or almost half-filled Hubbard model in the strong-coupling regime  is studied by the Kondo-lattice theory.
Physical properties of the Hubbard model, such as the self-energy, and the polarization function, and the three-point vertex function, are decomposed into their single-site and multisite properties.
Every single-site property can be mapped to its corresponding local property of the Anderson model. 
%
On the basis of  the mapping to the Anderson model, the Kondo temperature $T_{\rm K}$ or $k_{\rm B}T_{\rm K}$ is defined as 
the energy scale of local quantum spin fluctuations in the Hubbard model.   
The superexchange interaction can also be derived by field theory.
If the onsite repulsion is $U$ and the transfer integral between nearest neighbors is $-t/\sqrt{D}$, where $D$ is the dimensionality,
the exchange interaction constant between nearest neighbors is $J/D$, where $J=-4t^2/U$; this $J$ agrees with the one given by the conventional  derivation.
The Fock-type self-energy, which is of the first order in the superexchange interaction, is none other than the RVB self-energy.
Because of the RVB self-energy,
$k_{\rm B}T_{\rm K}$ is as large as $|J|/D$. 
If $D$ is sufficiently small such that no critical temperature $T_c$ exists or, even if $T_c$ exists, $T_c\ll T_{\rm K}$, 
then the RVB electron liquid is stabilized by the RVB self-energy at a sufficiently low $T$ such that $T\ll T_{\rm K}$ or $T_c<T\ll T_{\rm K}$.
If $U/|t|$ is finite or unless the filling of electrons is exactly half, $k_{\rm B}T_{\rm K}$ is nonzero and finite. 
Thus, the RVB electron liquid, which is characterized by $k_{\rm B}T_{\rm K}>0$, is stabler than any phase characterized by $k_{\rm B}T_{\rm K}=0$.
In the Heisenberg limit, the density of states at the chemical potential is almost vanishing, so that the RVB electron liquid is almost a spin liquid or a quasi spin liquid.
However, the quasi spin liquid shows a metallic conductivity at a sufficiently low temperature, provided that impurity scatterings are sufficiently weak and they are treated in the Born approximation.


According to the previous studies on the Kondo effect,
the local electron liquid in the Anderson model and the local spin liquid in the $s$-$d$ model are adiabatically connected to each other, although the local gauge symmetry does not exists in the Anderson model while it exists in the $s$-$d$ model.
According to the scaling theory for the Anderson localization,
if no symmetry breaking nor restoration occurs in a metal-insulator transition, there can be no critical point between metallic and insulating phases.
This fact means that  
the metallic and insulating phases are adiabatically connected to each other, even if the conductivity of the metallic phase is divergent and that of the insulating phase is zero. 
On the basis of these previous studies 
and the study in the present paper,
it is proposed that the RVB electron liquid in the Hubbard model and the RVB spin liquid in the Heisenberg model are adiabatically connected to each other, even if the local gauge symmetry does not exist in the Hubbard model and the conductivity of the RVB electron liquid is metallic while the local gauge symmetry exists in the Heisenberg model and the conductivity of the RVB spin liquid is zero.
The difference of the local gauge symmetry is irrelevant to the adiabatic continuation.

%

\appendix
\section{Sum Rule for the Self-Consistent $\tilde{\Delta}(\varepsilon)$}
\label{AppSumRule}
An analytic function $F(\varepsilon+i0)$ is defined by
\begin{align}\label{EqF1}
F(\varepsilon+i0) = \bigl[\varepsilon +\mu -\epsilon_d - \Sigma_\sigma(\varepsilon+i0) \bigr] - 1/R_\sigma(\varepsilon+i0).
\end{align}
According to the mapping condition of Eq.\hskip2pt(\ref{EqMapCond3}), it immediately follows that
\begin{align}\label{EqF2}
{\rm Im}\hskip1pt F(\varepsilon+i0)=
\tilde{\Delta}(\varepsilon).
\end{align} 
On the other hand, $\Delta\Sigma_\sigma(\varepsilon+i0,{\bm k})\rightarrow 0$ as $\varepsilon\rightarrow\pm\infty$; e.g.,  
$\Delta\Sigma_\sigma^{\rm (RVB)}(\varepsilon+i0,{\bm k})\rightarrow 0$ as $\varepsilon\rightarrow\pm\infty$, as shown in Eq.\hskip2pt(\ref{EqRVB-Vanish}) in \S\ref{SecRVB-Single}.
Then, it is straightforward to show that
\begin{align}\label{EqF3}
\lim_{\varepsilon\rightarrow \pm \infty}F(\varepsilon+i0) &=
-\biggl[\frac1{L}\sum_{\bm k}2\varphi_D^2({\bm k}) \biggr]\frac{2t^2}{\varepsilon} + O\biggl(\frac1{\varepsilon^2}\biggr)
=-\frac{2t^2}{\varepsilon}  + O\biggl(\frac1{\varepsilon^2}\biggr).
\end{align}
Since $F(\varepsilon+i0)$ is analytic in the upper-half complex plane,  according to Eqs.\hskip2pt(\ref{EqF2}) and (\ref{EqF3}),
\begin{align}\label{EqSumRuleA}
\int_{-\infty}^{+\infty} \hskip-10pt d\varepsilon \tilde{\Delta}(\varepsilon)
= 2\pi t^2 .
\end{align}
This is none other than the sum rule for the self-consistent $\tilde{\Delta}(\varepsilon)$.

\section{Theoretical Constraints for the Self-Consistent $1/\tilde{\phi}_1$}
\label{AppAppSumRule}
\subsection{Lower limit of the self-consistent $1/\tilde{\phi}_1$}
\label{AppAppSumRule1}

If $U/|t|\gg1$ and $n\simeq 1$, the density of states $\rho(\varepsilon)$ has a three-peak structure with the Gutzwiller band between the upper and lower Hubbard bands; their band centers are $\varepsilon\simeq 0$ and $\varepsilon\simeq \pm (U/2)$, resepctively. 
In general, if $\rho(\varepsilon)=-(1/\pi){\rm Im}R_\sigma(\varepsilon+i0)$ has a peak at an $\varepsilon$,
${\rm Re}\hskip1ptR_\sigma(\varepsilon+i0)$ becomes zero at the $\varepsilon$ or a little different $\varepsilon$ from the $\varepsilon$.
Thus, 
${\rm Re\hskip1pt}R_\sigma(\varepsilon+i0)=0$
for $\varepsilon$'s such as $\varepsilon=\epsilon_0\simeq 0$ and $\varepsilon=\epsilon_{\pm}\simeq \pm U/2$.
If ${\rm Re}\hskip1ptR_\sigma(\varepsilon+i0)=0$,
Eq.\hskip2pt(\ref{EqMapCondDelta}) becomes simple: 
%
\begin{align}\label{EqMapCondA}
\pi\rho(\varepsilon) \tilde{\Delta}(\varepsilon) = 1 - \pi\rho(\varepsilon) \bigl| {\rm Im}\Sigma_\sigma(\varepsilon+i0)\bigr|.
\end{align}

If $n=1$ or $n=1\pm 0^+$, the band center of the Gutzwiller band is at the chemical potential; thus, $\epsilon_0=0$ or $\epsilon_0=\pm 0^+$.
According to Eq.\hskip2pt(\ref{EqExpansionAM}), $|{\rm Im}\Sigma_\sigma(+i0)|=O\bigl(\tilde{\phi}_1k_{\rm B}T^2/T_{\rm K}\bigr)$; and according to Eq.\hskip2pt(\ref{EqRhoNum1}), $\rho(0)=\tilde{\rho}(0)=O\bigl[1/(\tilde{\phi}_1k_{\rm B}T_{\rm K})\bigr]$. Thus, according to Eq.\hskip2pt(\ref{EqMapCondA}),
\begin{align}\label{EqDeltaRho}
\pi \rho(0) \tilde{\Delta}(0)= 1-O\big[(T/T_{\rm K})^2\bigr].
\end{align}
If $n=1$ or $n=1\pm 0^+$, Eq.\hskip2pt(\ref{EqRVB-Q2}) is satisfied rather than Eq.\hskip2pt(\ref{EqRVB-Q1}).
According to Eqs.\hskip2pt(\ref{EqRhoNum3}) and (\ref{EqDeltaRho}), 
%
\begin{align}\label{EqPeakHeighgt}
\tilde{\Delta}(0) &=
O\bigl[(\tilde{\phi}_1 t^2 )/(DU)\bigr].
\end{align}
Since $k_{\rm B}T_{\rm K}=O\bigl[t^2/(DU)\bigr]$, as shown in Eq.\hskip2pt(\ref{EqTKNum3}),
Eq.\hskip2pt(\ref{EqPeakHeighgt}) is consistent with Eq.\hskip2pt(\ref{EqD-Pi}).
If $\tilde{\phi}_1$ is large, $\tilde{\Delta}(0)$ is large and $\tilde{\Delta}(\varepsilon)$ has a peak at $\varepsilon=0$.
Since the peak width is $O(k_{\rm B}T_{\rm K})$ or $O[t^2/(DU)]$ and the peak height is given by (\ref{EqPeakHeighgt}), it follows that 
\begin{align}\label{EqSumRuleMid}
\int_{-O(k_{\rm B}T_{\rm K})}^{+O(k_{\rm B}T_{\rm K})}\hskip-10pt
d \varepsilon \tilde{\Delta}(\varepsilon)
= O[t^2/(DU)] \times O\bigl[(\tilde{\phi}_1 t^2 )/(DU)\bigr]
= O\bigl[\tilde{\phi}_1 t^4/(DU)^2\bigr].
\end{align}
According to the sum rule of Eq.\hskip2pt(\ref{EqSumRuleA}), Eq.\hskip2pt(\ref{EqSumRuleMid}) has to be smaller than $2\pi t^2$. 
Thus, 
\begin{align}\label{EqUpperLimit}
\tilde{\phi}_1 \le  (1/c_{\phi}^\prime) (DU)^2/t^2,\quad
1/\tilde{\phi}_1 \ge c_{\phi}^\prime t^2/(DU)^2,
\end{align}
where $c_{\phi}^\prime>0$ and $c_{\phi}^\prime=O(1)$.
Even if $n=1$ or $n=1\pm 0^+$, there is a theoretical upper limit for the self-consistent $\tilde{\phi}_1$ or a theoretical lower limit for the inverse of it, $1/\tilde{\phi}_1$.


%

According to Gutzwiller's theory \cite{Gutzwiller1,Gutzwiller2,Gutzwiller3}, there is another theoretical lower limit for the self-consistent $1/\tilde{\phi}_1$, as shown in Eq.\hskip2pt(\ref{EqGutzwiller}). Thus, at least
\begin{align}
1/\tilde{\phi}_1 \ge \max\bigl[(c_{\phi}^\prime t^2/(DU)^2, c_{\rm g}|n-1|\bigr],
\end{align}
have to be satisfied.
If the RVB mechanism is considered, and if $U$ is finite, and even if $n=1$, the self-consistent $\tilde{\phi}_1$ cannot be divergent; and  $1/\tilde{\phi}_1$ cannot be zero. 
Thus, if $n=1$ or $n~1\pm 0^+$, the critical $U_c$, which is defined in \S\ref{SecSelf-energy}, exists and it is infinite; and if $n\ne 1$, it does not exist.

If $U/|t|\gg1$ and $n\simeq 1$, the peak height and the bandwidth of the upper and lower Hubbard band are
$\rho(\epsilon_{\pm}) = O(1/|t|)$ and
$W_{\rm H}=O(|t|)$,
%
respectively.
Since $\bigl| {\rm Im}\Sigma_\sigma(\epsilon_{\pm}+i0)\bigr|=O(|t|)$,
$\tilde{\Delta}(\epsilon_{\pm}) = O(|t|)$.
Then, it follows that
\begin{align}\label{EqA10}
\int_{\epsilon_{\pm}-2|t|}^{\epsilon_{\pm}+2|t|} \hskip-10pt 
d\varepsilon \tilde{\Delta}(\varepsilon)
\simeq \tilde{\Delta}(\epsilon_{\pm}) W_{\rm H}= O(t^2).
\end{align}
This is consistent with the sum rule of Eq.\hskip2pt(\ref{EqSumRuleA}).

\subsection{Asymptotic behavior of the self-consistent $1/\tilde{\phi}_1$}
\label{AppAppSumRule2}
The hybridization energy of the mapped Anderson model is given by
\begin{align}\label{EqDeltaApp1}
\tilde{\Delta}(\varepsilon) =
\frac{\pi}{\tilde{L}}\sum_{\bm k} \bigl|\tilde{V}_{\bm k}\bigr|^2\delta\bigl[\varepsilon+\tilde{\mu}-\tilde{E}_{c}({\bm k})\bigr],
\end{align}
where $\tilde{L}$ is the number of unit cells, $\tilde{V}_{\bm k}$ is the hybridization matrix between localized and conduction electrons, and $\tilde{E}_{c}({\bm k})$ is the dispersion relation of a conduction electron.
The Fermi surface is defined by $\tilde{\mu}= \tilde{E}_{c}({\bm k}_{\rm F})$.
Thus, Eq.\hskip2pt(\ref{EqFiniteDelta}) is a sufficient condition for the existence of the Fermi surface. 

Since the $\tilde{\Delta}(\varepsilon)$ defined in terms of $\tilde{V}_{\bm k}$ and $\tilde{E}_{c}({\bm k})$ is crucial for the Kondo effect, as shown in Eq.\hskip2pt(\ref{EqAnderson-G}), it can be assumed without the loss of generality that $\tilde{V}_{\bm k}$ is constant such that $\tilde{V}_{\bm k}=\tilde{V}$. Then
\begin{align}\label{EqDeltaApp2}
\tilde{\Delta}(\varepsilon) =
\pi |\tilde{V}|^2 \tilde{\rho}_c(\varepsilon), \quad
\tilde{\rho}_c(\varepsilon)=
\frac1{\tilde{L}}\sum_{\bm k} \delta\bigl[\varepsilon+\tilde{\mu}-\tilde{E}_{c}({\bm k})\bigr],
\end{align}
where
$\tilde{\rho}_c(\varepsilon)$ is the density of states of the conduction band.
According to the sum rule of Eq.\hskip2pt(\ref{EqSumRuleA}), it follows that $\pi |\tilde{V}|^2=2t^2$.

The half-filled Anderson model, in which $\tilde{n}=1$ or $\tilde{n}=1\pm 0^+$,  can be mapped to the $s$-$d$ model  in the $s$-$d$ or Heisenberg limit of $U/|t|\rightarrow +\infty$, with $J=-4t^2/U$ being kept constant; 
the exchange interaction constant is given by
\begin{align}
\tilde{J}_{s\mbox{-}d}= -4|\tilde{V}|^2/U 
= - (8/\pi)(t^2/U),
\end{align}
and the density of states of the conduction band is equal to the $\tilde{\rho}(\varepsilon)$  in the Anderson model. The dimensionless coupling constant, which is defined by
\begin{align}\label{EqG-s-d}
\tilde{g}(\varepsilon)= \tilde{J}_{s\mbox{-}d}\tilde{\rho}_c(\varepsilon)
=- [4/(\pi U)] \tilde{\Delta}(\varepsilon),
\end{align}
is relevant for the Kondo effect in the $s$-$d$ model.

If $\tilde{g}(\varepsilon)$ is constant as a function of $\varepsilon$, for example, in the most-divergent approximation \cite{abrikosov}, the Kondo temperature is given by 
\begin{align}
k_{\rm B}T_{\rm K} = \tilde{W} e^{-1/|\tilde{g}(0)|}, 
\end{align}
where $\tilde{W}$ is the half of the conduction bandwidth.
If $\tilde{g}(0)>0$, $k_{\rm B}T_{\rm K}$ is nonzero.
Since the energy dependence of $\tilde{g}(\varepsilon)$ is the same  as that of $\tilde{\Delta}(\varepsilon)$, as shown in Eq.\hskip2pt(\ref{EqG-s-d}), in the mapped $s$-$d$ model, 
the energy dependence of $\tilde{g}(\varepsilon)$ has to be seriously considered.
According to the scaling theory for the $s$-$d$ model \cite{poorman,wilsonKG},
high-energy processes substantially renormalize fixed-point or eventual low-energy properties but they can cause no symmetry breaking; and
the eventual low-energy properties play a crucial role in the quenching of the localized spin by the Kondo effect. Thus, whether the  eventual $k_{\rm B}T_{\rm K}$ is zero or nonzero depends on whether the bare $\tilde{g}(0)$ is zero or nonzero.  If $\tilde{g}(0)>0$, the eventual $k_{\rm B}T_{\rm K}$ is nonzero; and if $\tilde{g}(0)=0$, the eventual $k_{\rm B}T_{\rm K}$ is zero.

If Eqs.\hskip2pt(\ref{EqPeakHeighgt}) and (\ref{EqG-s-d}) are used, it follows that
\begin{align}
|\tilde{g}(0)|=
O\bigl[\tilde{\phi}_1t^2/\bigl(DU^2\bigr)\bigr].
\end{align} 
%
There are three possibilities fo the asymptotic behaviors of $|\tilde{g}(0)|$ and $1/\tilde{\phi}_1$ in the  Heisenberg limit:
\begin{subequations}
\begin{align}
\label{EqAsymp1}
\lim_{U/|t|\rightarrow +\infty}|\tilde{g}(0)|=0,
\hskip14pt
& \quad 
\lim_{U/|t|\rightarrow +\infty}1/\tilde{\phi}_1 
\gg t^2/(DU^2),
\\ \label{EqAsymp2}
0< \hskip2pt \lim_{U/|t|\rightarrow +\infty}|\tilde{g}(0)|<+\infty,
& \quad \lim_{U/|t|\rightarrow +\infty}1/\tilde{\phi}_1 
=O\bigl[t^2/(DU^2)\bigr],
\\ \label{EqAsymp3}
\lim_{U/|t|\rightarrow +\infty}|\tilde{g}(0)|=+\infty,
\hskip0pt
&\quad
\lim_{U/|t|\rightarrow +\infty}1/\tilde{\phi}_1 
\ll t^2/(DU^2).
\end{align}
\end{subequations}
Equation\hskip2pt(\ref{EqAsymp1}) is inconsistent with the self-consistent $k_{\rm B}T_{\rm K}$ being nonzero,
and Eq.\hskip2pt(\ref{EqAsymp3}) is inconsistent with Eq.\hskip2pt(\ref{EqUpperLimit}).
Thus, Eq.\hskip2pt(\ref{EqAsymp2}) has to be satisfied in a self-consistent solution for $n=1$ and $n=1\pm 0^+$.


\section{Proof of the Equality of Eq.~(\ref{EqWardtdtd*})}
\label{AppEq}
There is a useful relation between $\Xi_D$, which is defined by Eq.\hskip2pt(\ref{EqXiD}), and the static $\pi_{xx}(0)$, which is defined by Eq.\hskip2pt(\ref{EqLowerPiJD}), as studied below. In the presence of magnetic impurities, they are described as
\begin{align}\label{EqIntC1}
\Xi_D & =
-\frac1{\pi L} \sum_{{\bm k}}  
\varphi_D({\bm k})
\int_{-\infty}^{+\infty}\hskip-10pt d\epsilon 
f(\epsilon)
\hskip1pt{\rm Im}\hskip1pt \overline{g}_\sigma(\epsilon+i0,{\bm k})
\nonumber \\ &=
-\frac{\sqrt{D}}{\pi L} \sum_{{\bm k}} \cos(k_1a)
\int_{-\infty}^{+\infty}\hskip-10pt d\epsilon 
f(\epsilon)
\hskip1pt{\rm Im}\hskip1pt \overline{g}_\sigma(\epsilon+i0,{\bm k}),
\\ \label{EqStaticPi}
\pi_{xx}(0) &= 
\frac{2}{\pi L} \sum_{\bm k} \sin^2(k_1 a)
\int_{-\infty}^{+\infty} \hskip-10pt d\epsilon 
f(\epsilon) \bigl[{\rm Im}\hskip1pt \overline{g}_\sigma(\epsilon+i0,{\bm k})\bigr] 
\bigl[{\rm Re} \hskip1pt \overline{g}_\sigma(\epsilon+i0,{\bm k})\bigr].
\end{align}
%
Equation\hskip2pt(\ref{EqIntC1}) is also given, in the integration form, by
\begin{align}\label{EqIntC2}
\Xi_D &=
\frac{\sqrt{D}a^D}{\pi(2\pi)^D} \hskip-1pt
\hskip-2pt\int_{-\pi/a}^{+\pi/a} \hskip-15pt dk_1 
\cdots \hskip-3pt \int_{-\pi/a}^{+\pi/a} \hskip-15pt dk_D \hskip0pt
\cos(k_1a) 
\int_{-\infty}^{+\infty} \hskip-10pt d\epsilon f(\epsilon)
\hskip1pt{\rm Im}\hskip1pt \overline{g}_\sigma(\epsilon+i0,{\bm k}).
\end{align}
By the partial integration of Eq.~(\ref{EqIntC2}) with respect to $k_1$, it follows that 
\begin{align}\label{EqIntS1}
\Xi_D &= 
2t^*
\frac{a^D}{\pi (2\pi)^D} \hskip-2pt\int_{-\pi/a}^{+\pi/a} \hskip-15pt dk_1 \cdots \hskip-2pt\int_{-\pi/a}^{+\pi/a} \hskip-15pt dk_D\hskip2pt
\hskip-2pt\sin^2(k_1a)\hskip-2pt 
\int_{-\infty}^{+\infty} \hskip-12pt d\epsilon 
f(\varepsilon) 
\bigl[ {\rm Im}\hskip1pt \overline{g}_\sigma(\epsilon+i0,{\bm k})\bigr]
\bigl[ {\rm Re} \hskip1pt \overline{g}_\sigma(\epsilon+i0,{\bm k})\bigr].
\end{align}
This is also given, in the sum form, by
\begin{align}\label{EqIntS2}
\hskip-2pt
\Xi_D  &=
2t^*
\frac{2}{\pi L}\sum_{\bm k} \sin^2(k_1a)
\int_{-\infty}^{+\infty} \hskip-10pt d\epsilon 
f(\varepsilon) 
\bigl[ {\rm Im}\hskip1pt \overline{g}_\sigma(\epsilon+i0,{\bm k})\bigr]
\bigl[ {\rm Re} \hskip1pt \overline{g}_\sigma(\epsilon+i0,{\bm k})\bigr].
\end{align}
This is simply $2t^* \pi_{xx}(0)$, so that
\begin{align}\label{EqXiPi}
\Xi_{D}
= 2\left(t - \frac{3}{4}\tilde{\phi}_1 \tilde{W}_s^2 \Xi_{D}\frac{J}{D}\right)
\pi_{xx}(0).
\end{align}
Equation\hskip2pt(\ref{EqWardtdtd*}) is derived from Eq.\hskip2pt(\ref{EqXiPi}).


\begin{thebibliography}{99}
%
%
\bibitem{yosida}
K. Yosida{,}
Phys. Rev. {\bf 147}, 223 (1966).
%
\bibitem{poorman} 
P. W. Anderson{,}
J. Phys. C {\bf 3}, 2436 (1970).
%
\bibitem{wilsonKG} 
K. G. Wilson{,} 
Rev. Mod. Phys. {\bf 47}, 773 (1975). 
%
\bibitem{nozieres}
P. Nozi\`{e}res{,} 
J. Low. Temp. Phys. {\bf 17}, 31 (1974).
%
\bibitem{yamada1}
K. Yamada{,} 
Prog. Theor. Phys. {\bf 53}, 970 (1975).
%
\bibitem{yamada2}
K. Yamada and K. Yosida{,} 
Prog. Theor. Phys. {\bf 53}, 1286 (1975).
%
\bibitem{shiba}
H. Shiba{,}
Prog. Theor. Phys. {\bf 54}, 967 (1975).
%
%
%
%
%
\bibitem{hubbard1}
J. Hubbard, 
Proc. R. Soc. London Ser. A {\bf 276}, 238 (1963).
%
\bibitem{hubbard3}
J. Hubbard,
Proc. R. Soc. London Ser. A {\bf 281}, 401 (1964).
%
\bibitem{Gutzwiller1} 
M. C. Gutzwiller, 
Phys. Rev. Lett. {\bf 10}, 159 (1963).
%
\bibitem{Gutzwiller2} 
M. C. Gutzwiller, 
Phys. Rev. {\bf 134}, A923 (1964).
%
\bibitem{Gutzwiller3} 
M. C. Gutzwiller, 
Phys. Rev. {\bf 137}, A1726 (1965).
%
\bibitem{midband}
F. J. Ohkawa{,}
J. Phys. Soc. Jpn. {\bf 58}, 4156 (1989).
%
\bibitem{Js-mech-pert}
P. W. Anderson,
{\it Magnetism I}, ed. by G. T. Rado and H. Suhl,
(Academic Press, New York and London, 1963).
%
\bibitem{FJO-supJ}
F. J. Ohkawa{,}
Phys. Rev. B {\bf 65} (2002), 174424.
%
\bibitem{fazekas} 
P. Fazekas and P. W. Anderson,
Philos. Mag. {\bf 30}, 423 (1974).
%
\bibitem{highTc}
P. W. Anderson,
Science, {\bf 237}, 1196 (1987).
%
\bibitem{plain-vanilla}
P. W. Anderson, P. A. Lee, M. Randeria, T. M. Rice, N. Trivedi, and F. C. Zhang,
J. Phys. Condens. Matter {\bf 16}, R755 (2004),
%
\bibitem{Mapping-1}
F. J. Ohkawa{,} 
Phys. Rev. B {\bf 44}, 6812 (1991).
%
\bibitem{Mapping-2} 
F. J. Ohkawa{,} 
J. Phys. Soc. Jpn. {\bf 60}, 3218 (1991).
%
\bibitem{Mapping-3} 
F. J. Ohkawa{,} 
J. Phys. Soc. Jpn. {\bf 61}, 1615 (1992).
%
\bibitem{FJO-review}
F. J. Ohkawa,
J. Phys. Soc. jpn. {\bf 69} Suppl. A, 13 (2000).
%
\bibitem{toyama}
F. J. Ohkawa and T. Toyama{,}
J. Phys. Soc. Jpn. {\bf 78}, 124707 (2009).
%
\bibitem{FJO-MottIns}
F. J. Ohkawa{,} 
Porg. Theor. Phys. {\bf 128}, 125 (2012).
%
\bibitem{Metzner}
W. Metzner and D. Vollhardt{,}
Phys. Rev. Lett. {\bf 62}, 324 (1989).
%
\bibitem{Muller-H1}
E. M\"{u}ller-Hartmann{,}
Z. Phys. B {\bf 74}, 507 (1989).
%
\bibitem{Muller-H2}
E. M\"{u}ller-Hartmann{,}
Z. Phys. B {\bf 76}, 211 (1989).
%
\bibitem{Janis}
V. Janis{,}
Z. Phys. B {\bf 83}, 227 (1991). 
%
\bibitem{mermin}
N. D. Mermin and H. Wagner{,} 
Phys. Rev. Lett. {\bf 17}, 1133 (1966).
%
\bibitem{georges}
A. Georges and G. Kotliar{,} 
Phys. Rev. B {\bf 45}, 6479 (1992).
%
\bibitem{RevModDMFT}
A. Georges, G. Kotliar, W. Krauth and M. J. Rosenberg{,}
Rev, Mod. Phys. {\bf 68}, 13 (1996).
%
\bibitem{dcpa}
Y. Kakehashi and P. Fulde{,}
Phys. Rev. B {\bf 69}, 045101 (2004).
%
\bibitem{lieb-wu}
E. H. Lieb and F. Y. Wu{,} 
Phys. Rev. Lett. {\bf 20}, 1445 (1968). 
%
\bibitem{Luttinger1}
J. M. Luttinger and J. C. Ward,
Phys. Rev. {\bf 118}, 1417 (1960). 
%
\bibitem{Luttinger2}
J. M. Luttinger,
Phys. Rev. {\bf 119}, 1153 (1960). 
%
\bibitem{ward}
J. C. Ward,
Phys. Rev. {\bf 68}, 182 (1950).
%
\bibitem{miyai}
E. Miyai and F. J. Ohkawa,
Phys. Rev. B {\bf 61}, 1357 (2000).
%
\bibitem{satoh}
H. Satoh and F. J. Ohkawa,
Phys. Rev. B {\bf 63}, 184401 (2001).
%
\bibitem{kubo}
R. Kubo,
J. Phys. Soc. Jpn. {\bf 12}, 570 (1957).
%
\bibitem{elitzur}
S. Elitzur,
Phys. Rv. D {\bf 12}, 3978 (1975).
%
\bibitem{abrahams}
E. Abrahams, P. W. Anderson, D. C. Licciardello, and T. V. Ramakrishnan,
Phys. Rev. Lett. {\bf 42}, 673 (1979).
%
\bibitem{solyom}
J. S$\hat{\rm o}$lyom,
Advances in Phys. {\bf 28}, 201 (1979).
%
\bibitem{Luttinger-liq}
J. M. Luttinger, 
J. Math. Phys. {\bf 4}, 1154 (1963).
%
\bibitem{cloiseaux}
J. des Cloizeaux and J. J. Pearson,
Phys. Rev. {\bf 128}, 2131 (1962).
%
\bibitem{TL1}
R. Chitra and T. Giamarch,
Phys. Rev. B {\bf 55}, 5816 (1997).
%
\bibitem{TL2}
T. Giamarch and A. M. Tsvelik,
Phys. Rev. B {\bf 59}, 11398 (1999).
%
\bibitem{TL3}
A. Furusaki and F. C. Zhang,
Phys. Rev. B {\bf 60}, 1175 (1999).
%
%
\bibitem{HZ2-1}
A. Auerbach and D. P Arovas,
Phys. Rev. Lette. {\bf 61}, 617 (1988).
%
\bibitem{HZ2-2}
S. Miyashita,
J. Phys. Soc. Jpn. {\bf 57}, 1934 (1988).
%
\bibitem{slater}
J. C. Slater{,}
Phys. Rev. {\bf 82}, 538 (1951).
%
\bibitem{Anderson-SC}
P. W. Anderson{,} 
Science {\bf 235}, 1196 (1987).
%
\bibitem{FJO-SC1}
F. J. Ohkawa{,}
J. Phys. Soc. Jpn. {\bf 56}, 2267 (1987).
%
\bibitem{FJO-SC2}
F. J. Ohkawa{,}
J. Phys. Soc. Jpn. {\bf 78}, 084712 (2009).
%
\bibitem{abrikosov}
A. A. Abrikosov,
Physics {\bf 2}, 5 (1965).
%
\end{thebibliography}
\end{document}